\documentclass[floatfix,twocolumn,astrosymb,tighten]{aastex63}
\usepackage[utf8]{inputenc}
\usepackage{amsmath}
\usepackage{amsmath}
\usepackage{pgfplots}
\usepackage{amsmath}
\usepackage{multirow}
\usepackage{amsmath}
\usepackage[maxfloats=256]{morefloats}
\usepackage{hyperref}
\usepackage{tikz}
\usepackage{gensymb}
\pgfplotsset{compat=1.16}

\graphicspath{{./}{figures/}}

\begin{document}

\title{A candidate relativistic tidal disruption event at 340 Mpc}

\author[0000-0001-8426-5732]{Jean J. Somalwar}
\affil{Cahill Center for Astronomy and Astrophysics, MC\,249-17 California Institute of Technology, Pasadena CA 91125, USA.}

\author[0000-0002-7252-5485]{Vikram Ravi}
\affil{Cahill Center for Astronomy and Astrophysics, MC\,249-17 California Institute of Technology, Pasadena CA 91125, USA.}

\author[0000-0001-9584-2531]{Dillon Z. Dong}
\affil{Cahill Center for Astronomy and Astrophysics, MC\,249-17 California Institute of Technology, Pasadena CA 91125, USA.}

\author{Yuyang Chen}
\affil{David A. Dunlap Department of Astronomy and Astrophysics, University of Toronto, 50 St. George Street, Toronto M5S 3H4, Canada}
\affil{Dunlap Institute for Astronomy and Astrophysics, University of Toronto, 50 St. George Street, Toronto M5S 3H4, Canada}

\author{Shari Breen}
\affil{SKA Observatory, Jodrell Bank, Lower Withington, Macclesfield, SK11 9FT, UK}

\author[0000-0002-0844-6563]{Poonam Chandra}
\affil{National Radio Astronomy Observatory, 520 Edgemont Road, Charlottesville VA 22903, USA}

\author[0000-0001-6812-7938]{Tracy Clarke}
\affil{Naval Research Laboratory, 4555 Overlook Ave SW, Washington, DC 20375, USA.}

\author{Kishalay De}
\affiliation{MIT-Kavli Institute for Astrophysics and Space Research, 77 Massachusetts Ave., Cambridge, MA 02139, USA}

\author[0000-0002-3382-9558]{B. M. Gaensler}
\affil{Dunlap Institute for Astronomy and Astrophysics, University of Toronto, 50 St. George Street, Toronto M5S 3H4, Canada}
\affil{David A. Dunlap Department of Astronomy and Astrophysics, University of Toronto, 50 St. George Street, Toronto M5S 3H4, Canada}

\author[0000-0002-7083-4049]{Gregg Hallinan}
\affil{Cahill Center for Astronomy and Astrophysics, MC\,249-17 California Institute of Technology, Pasadena CA 91125, USA.}

\author[0000-0003-2714-0487]{Sibasish Laha}

\affiliation{Center for Space Science and Technology, University of Maryland Baltimore County, 1000 Hilltop Circle, Baltimore, MD 21250, USA.}
\affiliation{Astrophysics Science Division, NASA Goddard Space Flight Center, Greenbelt, MD 20771, USA.}
\affiliation{Center for Research and Exploration in Space Science and Technology, NASA/GSFC, Greenbelt, Maryland 20771, USA}

\author[0000-0002-4119-9963]{Casey Law}
\affil{Cahill Center for Astronomy and Astrophysics, MC\,249-17 California Institute of Technology, Pasadena CA 91125, USA.}

\author{Steven T. Myers}
\affil{National Radio Astronomy Observatory, P.O. Box O, Socorro, NM 87801, USA}

\author{Tyler Parsotan}
\affil{Center for Space Science and Technology, University of Maryland Baltimore County, 1000 Hilltop Circle, Baltimore, MD 21250, USA.}
\affil{Astrophysics Science Division, NASA Goddard Space Flight Center, Greenbelt, MD 20771, USA.}
\affil{Center for Research and Exploration in Space Science and Technology, NASA/GSFC, Greenbelt, Maryland 20771, USA}

\author[0000-0002-5187-7107]{Wendy Peters}
\affil{Naval Research Laboratory, 4555 Overlook Ave SW, Washington, DC 20375, USA.}

\author{Emil Polisensky}
\affil{Naval Research Laboratory, 4555 Overlook Ave SW, Washington, DC 20375, USA.}

\begin{abstract}
We present observations of an extreme radio flare, VT J024345.70-284040.08, hereafter VT J0243, from the nucleus of a galaxy with evidence for historic Seyfert activity at redshift $z=0.074$. Between NRAO VLA Sky Survey observations in 1993 to VLA Sky Survey observations in 2018, VT J0243 rose from a ${\sim}$GHz radio luminosity of $\nu L_\nu \lesssim 10^{38}$ erg s$^{-1}$ to $\nu L_\nu{\sim}10^{40}$ erg s$^{-1}$, and still continues to brighten. The radio spectral energy distribution (SED) evolution is consistent with a nascent jet that has slowed over ${\sim}3000$ days with an average $0.1 < \langle \beta \rangle < 0.6$. The jet is energetic (${\sim}10^{51-52}$ erg), and had a radius ${\sim}0.7$ pc in Dec. 2021. X-ray observations suggest a persistent or evolving corona, possibly associated with an accretion disk, and IR and optical observations constrain any high-energy counterpart to be sub-Eddington. VT J0243 may be an example of a young, off-axis radio jet from a slowly evolving tidal disruption event. Other more mysterious triggers for the accretion enhancement and jet launching are possible. In either case, VT J0243 is a unique example of a nascent jet, highlighting the unknown connection between supermassive black holes, the properties of their accretion flows, and jet launching.
\end{abstract}

\section{Introduction} \label{sec:intro}

In galactic nuclei, accretion-associated phenomena produce extreme radio variability on timescales of months$-$years and with flare luminosities covering the full range up to ${\gtrsim}10^{32}$ erg s$^{-1}$ Hz$^{-1}$. These flares are often associated with tidal disruption events (TDEs; \citealp{Alexander2020}), young radio jets from active galactic nuclei (AGN; \citealp{Nyland2020, Mooley2016, Kunert2020, Wolowska2021}), or outflows launched from accretion disks \citep[e.g.][]{Koay2016}. 

The physical mechanisms causing many of these radio flares in galactic nuclei have yet to be understood. For stellar mass black holes, it is well established that there is a strong connection between accretion and the launching of radio-emitting outflows and jets \citep[see][for a review]{Fender2010}. Jet and outflow launching from supermassive black holes (SMBHs) is an unsolved problem, whether we consider those black holes associated with AGN that have been accreting for long timescales or the newly active black holes resulting from stellar disruptions. The conditions under which radio jets launch, the mechanisms through which they emit across the electromagnetic spectrum, and their duty cycle remain open questions \citep[e.g.][]{Blandford2019}.

Our understanding of accretion-associated radio transients is evolving significantly with the advent of high-resolution, wide-field radio surveys, such as the Very Large Array Sky Survey (VLASS; \citealp{Lacy2020}). In this paper, we present an extraordinarily luminous radio transient discovered with VLASS, VT J024345.70-284040.08, hereafter VT J0243. VT J0243 is located in the nucleus of a nearby galaxy, 2dFGRS TGS314Z138 ($z=0.0742$, Section~\ref{sec:host}). We identified VT J0243 as a radio transient between the NRAO VLA Sky Survey (NVSS; \citealp{Condon1998}) and VLASS.  VT J0243 has risen to ${\sim}10^{40}$ erg s$^{-1}$ over ${\sim}5$ years, and continues to brighten. As we will show, VT J0243 is likely a nascent radio jet, yet no other event observed to date has shown its multiwavelength signatures, including a radio luminosity that continues to rise more than $1000$ days after the jet turned on. In Section~\ref{sec:targetsel}, we describe our selection criteria that led to the discovery of VT J0243. In Section~\ref{sec:obs}, we describe our multiwavelength archival searches and follow-up observations. In Section~\ref{sec:ana}, we present our analysis of the observations, and in Section~\ref{sec:disc}, we discuss the interpretation of VT J0243. 

We adopt the \cite{Planck2020} cosmology with $H_0 = 67.7$ km s$^{-1}$ Mpc$^{-1}$.

\section{Target Selection} \label{sec:targetsel}

VT J0243 was detected as part of our transient search using the 1.4 GHz NRAO VLA Sky Survey in the 1990s \citep[][]{Condon1998} and the 3 GHz VLA Sky Survey \citep[][]{Lacy2020}, observing from 2017 to today. These surveys provide a unique opportunity to identify slowly evolving radio transients. NVSS has an rms noise 0.45 mJy/beam and a resolution of $45\arcsec$ FWHM, and VLASS has an rms noise 0.14 mJy/beam and a resolution 1\farcs5 FWHM. Dong et al., in prep., generated a transient catalog by identifying sources that were detected by \texttt{pyBDSF} at a $>7\sigma$ level in VLASS but were not detected ($<3\sigma$) in NVSS. We refer the reader to that work and Appendix A of \cite{Somalwar2021} for a detailed description of the pipeline used.

VT J0243 was also selected as an evolving source in an independent search (Chen et al., in prep.) that identified young radio transients through VLASS and the VLITE Commensal Sky Survey \citep[VCSS;][]{peters2021}. VCSS is a survey conducted simultaneously with VLASS by VLITE, a commensal instrument on the VLA \citep{clarke2016,polisensky2016}. VCSS covers the same regions of the sky as VLASS and observes at $\nu\sim340\,\mathrm{MHz}$ with an angular resolution of $\theta\sim20''$ and a median image rms of $3\,\mathrm{mJy/beam}$. 
Additionally, VT J0243 was identified to be young because of its inverted spectrum between $340\,\mathrm{MHz}-3\,\mathrm{GHz}$, suggesting optically thick emission at low frequencies.

Because of the extreme radio luminosity of this source given its history of inactivity and its coincidence with the nucleus of a low-mass galaxy, we initiated an extensive, multi-wavelength follow-up campaign.

\section{Observations and Data Reduction} \label{sec:obs}

\begin{deluxetable*}{ccccc}
\tablecaption{Radio Observations  \label{tab:radio}}
\tablewidth{0pt}
\tablehead{\colhead{Instrument/Survey} & \colhead{Date} & \colhead{MJD} & \colhead{Frequency [GHz]} & \colhead{Flux Density [mJy]} }
\startdata
NVSS$^{(1)}$ & Sept. 20 1993 & 49250 & 1.4 & $< 1.3$ ($3\sigma$) \\\hline
TGSS$^{(2)}$ & Dec. 27 2010 & 55557 & 0.15 & $< 15$ ($3\sigma$) \\\hline
VCSS$^{(3)}$ Epoch 1 & Feb. 17, 2018 & 58166 & 0.340 & $23 \pm 7$ \\\hline
VLASS$^{(4)}$ Epoch 1 & Feb. 17, 2018 & 58166 & 2.157 & $39.14 \pm 0.31$  \\\hline
VLASS Epoch 1 & Feb. 17, 2018 & 58166 & 2.578 & $40.21 \pm 0.26$  \\\hline
VLASS Epoch 1 & Feb. 17, 2018 & 58166 & 3.048 & $41.54 \pm 0.30$  \\\hline
VLASS Epoch 1 & Feb. 17, 2018 & 58166 & 3.865 & $41.39 \pm 0.34$  \\\hline
RACS$^{(5)}$ & Apr. 28, 2019 & 58601 & 0.8875 &  $45.81 \pm 0.64$ \\\hline
VCSS Epoch 2 & Nov. 1, 2020 & 59154 & 0.340 & $22 \pm 7$  \\\hline
VLASS Epoch 2 & Nov. 1, 2020 & 59154 & 2.157 & $54.33 \pm 0.48$ \\\hline
VLASS Epoch 2 & Nov. 1, 2020 & 59154 & 2.579 & $53.85 \pm 0.34$ \\\hline
VLASS Epoch 2 & Nov. 1, 2020 & 59154 & 3.048 & $52.84 \pm 0.29$ \\\hline
VLASS Epoch 2 & Nov. 1, 2020 & 59154 & 3.685 & $52.29 \pm 0.34$ \\\hline
ATCA Epoch 1 (PC: CX486) & Jun. 27, 2021 & 59392 & 1.877 & $68.82 \pm 11.51$ \\\hline
ATCA Epoch 1 (PC: CX486) & Jun. 27, 2021 & 59392 & 2.636 & $71.26 \pm 12.69$ \\\hline
ATCA Epoch 1 (PC: CX486) & Jun. 27, 2021 & 59392 & 4.79 & $59.23 \pm 6.59$ \\\hline
ATCA Epoch 1 (PC: CX486) & Jun. 27, 2021 & 59392 & 5.779 & $55.92 \pm 6.24$ \\\hline
ATCA Epoch 1 (PC: CX486) & Jun. 27, 2021 & 59392 & 6.732 & $52.48 \pm 5.34$ \\\hline
ATCA Epoch 1 (PC: CX486) & Jun. 27, 2021 & 59392 & 7.734 & $49.27 \pm 5.22$ \\\hline
ATCA Epoch 1 (PC: CX486) & Jun. 27, 2021 & 59392 & 8.706 & $46.49 \pm 4.9$ \\\hline
ATCA Epoch 1 (PC: CX486) & Jun. 27, 2021 & 59392 & 9.677 & $45.08 \pm 4.98$ \\\hline
ATCA Epoch 1 (PC: CX486) & Jun. 27, 2021 & 59392 & 10.68 & $43.61 \pm 5.56$ \\\hline
ATCA Epoch 2 (PC: CX486) & Aug. 13, 2021 & 59439 & 5.25 & $57.47 \pm 0.7$ \\\hline
ATCA Epoch 2 (PC: CX486) & Aug. 13, 2021 & 59439 & 8.75 & $48.74 \pm 0.52$ \\\hline
ATCA Epoch 2 (PC: CX486) & Aug. 13, 2021 & 59439 & 18.0 & $32.0 \pm 0.66$ \\\hline
ATCA Epoch 2 (PC: CX486) & Aug. 13, 2021 & 59439 & 34.0 & $21.4 \pm 0.71$ \\\hline
ATCA Epoch 2 (PC: CX486) & Aug. 13, 2021 & 59439 & 40.0 & $19.82 \pm 0.69$ \\\hline
GMRT (PID: ddtC203) & Aug. 27, 2021 & 59454 & 0.402 & $ 32.39 \pm 0.24 $  \\\hline
GMRT (PID: ddtC203) & Aug. 29, 2021 & 59454 & 0.648 & $ 52.44 \pm 0.26 $  \\\hline
GMRT (PID: ddtC203) & Aug. 28, 2021 & 59454 & 1.264 & $ 60.88 \pm 0.41 $  \\\hline
\enddata
\tablecomments{Archival and follow-up radio observations of VT J0243. References: $^{(1)}$\citep[][]{Condon1998}, $^{(2)}$\citep[][]{Intema2017}, $^{(3)}$\citep[][]{peters2021}, $^{(4)}$\citep[][]{Lacy2020}, $^{(5)}$\citep[][]{McConnell2020}.}
\end{deluxetable*}

In this section, we describe our multi-wavelength follow-up of and archival searches for VT J0243 and its host, 2dFGRS TGS314Z138. 

\subsection{Radio observations} \label{sec:radioobs}

The available archival radio observations and our radio follow-up are summarized in Table~\ref{tab:radio}. After a nondetection by NVSS on MJD 49520, VT J0243 was first detected on MJD 58166 in the first epoch of the VLASS with a luminosity $\nu L_{\nu}({\rm 3\,GHz}) \sim 10^{40}$ erg s$^{-1}$. NVSS and VLASS are described at the beginning of Section~\ref{sec:targetsel}. At the same time as the VLASS first epoch observations, VCSS detected the source (see Section~\ref{sec:targetsel} for details of VCSS). The source was then detected by the Autralian SKA Pathfinder (ASKAP) telescope as part of the Rapid ASKAP Continuum Survey (RACS) at 0.9 GHz \cite{McConnell2020}. RACS is observing the whole sky visible to ASKAP in the $700{-}1800$ MHz band with 15\arcsec resolution and a sensitivity of ${\sim}0.25$ mJy/beam. The final surveys to detect VT J0243 were the second epochs of VLASS and VCSS. Follow-up observations for this source were obtained using the Australia Telescope Compact Array (ATCA), the upgraded Giant Metrewave Radio Telescope (uGMRT), and the Very Long Baseline Array (VLBA).

Two epochs of ATCA observations were obtained on MJDs 59392 and 59439 with the six 22\,m dishes arranged in the extended 6B configuration, providing baselines spanning 214--5969\,m.\footnote{\url{https://www.narrabri.atnf.csiro.au/observing/users_guide/html/chunked/aph.html}}. The Compact Array Broadband Backend \citep[CABB;][]{wilson11} was used in the \texttt{CFB-1M} mode to simultaneously record full-polarization visibilities in two 2048\,MHz bands each split into 2048 1\,MHz channels. In the first epoch, by cycling between three different non-standard frequency setups data were obtained in 2048\,MHz bands centered on 2.1\,GHz, 5.25\,GHz, 7\,GHz, 8.75\,GHz, and 10.25\,GHz. Observations in the first epoch totaled two hours. Scans of PKS\,1934$-$638 in each frequency setup were used to set the flux-density scale, and calibrate the complex time-independent bandpasses. Regular observations of the unresolved source PKS\,0237$-$233 were used to calibrate the time-variable complex gains. In the second epoch, data were obtained at 5.25\,GHz, 8.75\,GHz, 18\,GHz, 24\,GHz, 34\,GHz, 40\,GHz in 2048\,MHz bands to further constrain time evolution and spectral shape at high frequencies. For the cm bands, scans of PKS 1934-638 and PKS 0237-233 were again used to calibrate the bandpass, flux density scale, and time-variable gains. For the mm bands, scans of PKS 1921-293 were used instead of PKS 1934-638 for the bandpass and flux calibration.

The data were reduced, edited, calibrated and imaged using standard techniques implemented in the MIRIAD package \citep{miriad}. Multi-frequency synthesis images were made in multiple sub-bands, centered on frequencies listed in Table~\ref{tab:radio}. VT J0243 was detected in all images; single rounds of phase-only self calibration were applied in each band to improve image quality. Flux densities and their uncertainties were estimated using the MIRIAD task \texttt{imfit}. 

The event VT J0243 was observed with the upgraded Giant Metrewave Radio Telescope (uGMRT) under Director’s Discretionary Time (DDT) proposal DDT C203 on 2021 Aug 27, 28 and 29 in bands 3 (250—500 MHz),  5 (1000– 1450 MHz) and 4 (550—900 MHz), respectively, of the uGMRT. The observations were two hours in duration including overheads using a bandwidth of 400 MHz in bands 4 and 5, whereas the duration was three hours in band 3. The VLA calibrator 3C 147 was used as a flux and a bandpass calibrator and J0240-231 was used as a phase calibrator. We use the Common Astronomy Software Applications (CASA; \citealp{McMullin2007}) for data analysis. The data were analyzed in three major steps, i.e flagging, calibration and imaging using the procedure laid out in \cite{Maity2021}. A total of 6 rounds of phase self-calibrations and 2 rounds of amplitude \& phase self-calibration  were performed. A source was clearly detected at the VLASS position. The source flux densities  at  bands 5, 4 and 3 are mentioned in Table~\ref{tab:radio}. 

VLBA observations of VT J0243 were conducted on MJD 59569, with 512\,MHz of bandwidth centered on 8.368\,GHz, and the data were processed using the DiFX correlator \citep{deller11}. Data were recorded at a rate of 4.096\,Gbps at all sites besides North Liberty in four 128\,MHz sub-bands, using the Digital Downconverter (DDC) mode of the Roach Digital Backends. Given the high expected flux density of the source, we planned to self-calibrate the observations. The 45\,min observation included two 2\,min scans of the fringe finder J0555+3948, and two 1\,min scans of the check source J0236-2953, and a total of 31.5\,min on VT J0243. Calibration and imaging of the observations was carried out using CASA, following procedures outlined in VLBA Memo 38.\footnote{\url{https://library.nrao.edu/public/memos/vlba/sci/VLBAS_38.pdf}} Following data editing, we performed a global fringe-fit, which was successful for seven antennas (data from Pie Town and St. Croix were substantially lower in sensitivity). We then performed two rounds of phase-only self-calibration on VT J0243, and one round of amplitude$+$phase self-calibration. This yielded phase variations under $\pm5$\,deg. An image of and inspection of visibility amplitudes on VT J0243 revealed a partially resolved source. We fit the data with an elliptical Gaussian model using the CASA task \texttt{uvmodelfit}, and found a flux density of 37\,mJy (with $\sim10\%$ uncertainty), a major axis of $1.1\pm0.1$\,mas, and a minor axis of $0.5\pm0.1$\,mas, at a position angle of $-23$\,deg.  

\subsection{Optical photometry}
From the radio observations of VT J0243, we can naively constrain the radio-turn-on time range to $1990-2018$. The Catalina Realtime Transient Survey (CRTS; \citealp{Drake2009}) observed the location of VT J0243 between ${\sim}2005$ and $2013$ (MJD $53554-56302$), the Pan-STARRS $3\pi$ survey \citep{Chambers2016} over ${\sim}2010-2013$ (MJD $55433-56970$), 
and the Asteroid Terrestrial-impact Last Alert System (ATLAS; \citealp{Tonry2018}) between ${\sim}2015-2021$ (MJD $57303-59097$). We retrieve the CRTS photometry for this source from the default \texttt{photcat} catalog \citep[][]{Drake2009}. This photometry is performed on absolute (i.e., not difference) images using \texttt{SExtractor} to measure aperture magnitudes. Also note that CRTS does not use a filter, so the absolute calibration of the photometry is uncertain. We retrieved archival optical images of the source from the PanSTARRS1 survey \citep{Chambers2016}. The reduced images were processed through a custom image subtraction pipeline (described in \citealt{De2020}) to remove the host galaxy light using the first epoch of PS1 observations as a template. Point-spread function photometry was performed on the resulting difference images to derive the optical light curve shown in Figure~\ref{fig:lc}. We retrieve ATLAS photometry at the position of 2dFGRS TGS314Z138 from their forced photometry server\footnote{\url{https://fallingstar-data.com/forcedphot/}} using default settings. Finally, we generated a mid-infrared lightcurve for VT J0243 by performing PSF photometry on single-epoch difference images from the UNWISE reprocessing of observations from the WISE and NEOWISE surveys \citep[][]{Lang2014, Meisner2017, Wright2010, Mainzer2011}. The resulting lightcurves are summarized in Figure~\ref{fig:lc}.

\subsection{Optical spectroscopy}
An optical spectrum of VT J0243 was obtained before 2002 (MJD $<52375$) as part of the 2dF Galaxy Redshift Survey (2dFGRS; \citealp{Cole2005}). The spectrum was taken using a 2\farcs0 arcsec fiber with the 2dF instrument on the Anglo-Australian Telescope telescope. The wavelength range was $3627-8037\,{\rm \AA}$ (observed frame) and the resolution $R = 648$. We retrieved the non-flux calibrated spectrum from the NASA/IPAC Extragalactic Database (NED). We observed VT J0243 on the night of Oct. 6 2021 (MJD 59493) using the Low Resolution Image Spectrometer (LRIS; \citealp{Oke1995}) on the Keck I telescope. We used the 1\farcs0 slit centered on the galactic nucleus using a parallactic angle ($-0.035^{\circ}$). We used the 400/3400 grism, the 400/8500 grating with central wavelength 7830, and the 560 dichroic. We observed this source for 20 min. The resulting wavelength range was ${\sim}1300{-}10000\,{\rm \AA}$ and the resolution $R{\sim}700$. Comparing these spectra, there are no obvious transient features (Figure~\ref{fig:lc}). Weak AGN-like emission lines are visible, but no broad emission lines are detected. The spectra are all fully consistent with being host-dominated.

\subsection{X-ray/UV observations}
VT J0243 was observed in the X-ray band as part of the ROSAT survey on Jan. 6, 1990 (MJD 47897). There is no detection reported in the Second ROSAT All-Sky Survey Point Source Catalog \citep[][]{Voges1993, Boller2016}. 
We retrieved the ROSAT image at the location of VT J02438 from the HEASARC archive\footnote{\url{https://heasarc.gsfc.nasa.gov/docs/archive.html}}, and used \texttt{ximage} to find a $3\sigma$ upper limit of the $0.3{-}10$ keV, unabsorbed soft X-ray flux, $f_X \lesssim 2\times10^{-13}$ erg cm$^{-2}$ s$^{-1}$, assuming a power law spectrum with $\Gamma=3$ and the Milky Way $N_{H, {\rm MW}} = 1.51 \times 10^{20}$ cm$^{-2}$ \citep[][]{HI4PI2016}. 
VT J0243 was subsequently observed by the {\it XMM}-Newton Slew Survey on Jul. 30, 2008 (MJD 54677), Dec. 30, 2009 (MJD 55195), Jul. 12, 2012 (MJD 56120), Jan. 26, 2021 (MJD 59240), and Jun. 26, 2021 (MJD 59391). No detection was reported on any of these dates, so we adopt an upper limit corresponding to the flux limit reported by the survey $f({\rm 0.2-12\,keV}) = 1.3 \times 10^{-12}$ erg s$^{-1}$ cm$^{-2}$ ($L \sim 1.67\times10^{43}$ erg s$^{-1}$). Finally, VT J0243 was observed by the Monitor of All-sky X-ray Image (MAXI; \citealp{Matsuoka2009}). We retrieved $2-30$ keV photometry for this transient using the on-demand photometry survey provided by the MAXI collaboration\footnote{\url{http://maxi.riken.jp/mxondem/}} for the MJD range $55058.0{-}58000.0$. No significant detection was reported, and the typical upper limit was $f_X({\rm 0.3-10\,keV}) \lesssim 7.7\times10^{-11}$ erg s$^{-1}$ cm$^{-2}$. There was also no significant detection by the Swift Burst Area Telescope (BAT) in the transient monitor light curve produced for this source during the MJD range $55798{-}56961$ \citep[][]{Krimm2006}.

We observed VT J0243 using the X-ray Telescope (XRT) on The Neil Gehrels Swift Observatory (Swift XRT; \citealp{Burrows2005}) on MJD 59379 and 59484 for 3 and 5 ks exposures, respectively. A source was detected in both exposures, with $0.3-10$ keV fluxes of $0.67\pm0.35 \times10^{-13}$ erg cm$^{-2}$ s$^{-1}$ and $1.28\pm0.37$ cm$^{-2}$ s$^{-1}$, respectively, assuming power-law spectra with $\Gamma=3$. We then obtained a soft X-ray spectrum for VT J0243 on MJD 59391 using the {\it XMM}-Newton observatory EPIC camera using the thin filters in full frame mode with a 30 ks exposure time. We used the standard analysis pipeline to process the data and extract an X-ray spectrum. 

Swift/UVOT observed VT J0243 simultaneously with the Swift/XRT observations in the UVW1 band. We reduced the observations using the standard HEASOFT pipeline and measured the source magnitude using the \texttt{uvotsource} tool with a 5\arcsec source region and a 15\arcsec background region offset from the source. We found a UVW1 AB magnitude $20.38 \pm 0.06 ({\rm stat}) \pm 0.03 ({\rm sys})$. This is consistent within $2\sigma$ with the quiescent-level predictions from our SED fit (see Section~\ref{sec:host}), so there is no detectable transient emission and we do not consider these observations further.

\section{Analysis} \label{sec:ana}

\begin{figure*}[t!]
\gridline{\fig{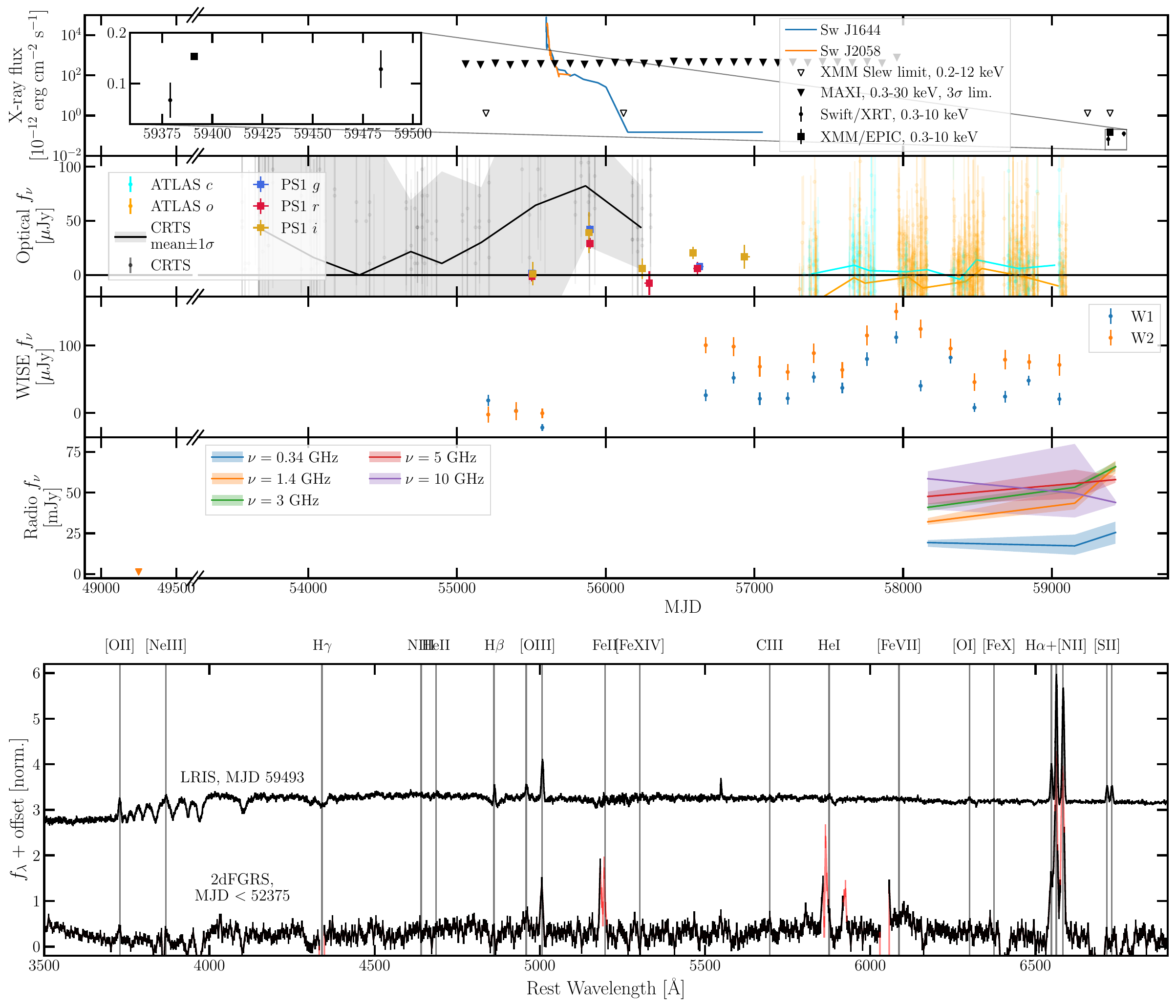}{\textwidth}{}}
\caption{ ({\it top}) From top to bottom, X-ray, optical, mid-infrared, and radio lightcurves for VT J0243. In the topmost panel, the black triangles represent three sigma upper limits from the MAXI and XMM slew surveys. The squares and circles show recent detections by Swift/XRT and XMM/EPIC, and the inset axis magnifies this data. The colored lines show the X-ray lightcurves for two of the three X-ray detected jetted TDEs, Sw J1644+57 \citep[][]{Levan2016} and Sw J2058+05 \citep[][]{Pasham2015}. We do not show the X-ray lightcurve for the final X-ray detected jetted TDE, Sw J1112-8283 \citep[][]{Brown2017}, as it largely overlaps with the Sw J1644+57 lightcurve but is poorly sampled in comparison. Regardless of the binning of the MAXI observations, an X-ray flare of the same luminosity as those detected for previous jetted TDEs would have been detected in the first ${\sim}100$ days. An optical flare is detected in PanSTARRS around MJD 56000. The MIR emission is measured using forced photometry on NEOWISE images, and is variable. The MIR color appears to have reddened after the first three epochs. The radio lightcurves are extrapolated from the model fits described in Section~\ref{sec:radio}, and the non-detection on the far left of the plot corresponds to the 1990 NVSS observation. ({\it bottom}) LRIS ({\it top}) and 2dFGRS ({\it bottom}) optical spectra, normalized. No significant transient features are detected. No broad emission lines are significantly detected. The emission line ratios are consistent with weak Seyfert emission on a BPT diagram. \label{fig:lc}
}
\end{figure*}

In this section, we present our analyses of the archival and follow-up observations of VT J0243. We begin in Section~\ref{sec:host} with a brief discussion of the host properties. In Section~\ref{sec:radio}, we constrain the physical properties of the radio-emitting outflow or jet using the radio observations. In Section~\ref{sec:xray} we constrain the origins of the X-ray emission, and in Section~\ref{sec:OIR}, we discuss the optical and infrared photometry at the location of VT J0243.

\subsection{Host Galaxy} \label{sec:host}

\begin{deluxetable}{cc}
\tablecaption{Host Galaxy, 2dFGRS TGS314Z138  \label{tab:host}}
\tablewidth{0pt}
\tablehead{\colhead{Parameter} & \colhead{Value} }
\startdata
R.A. (J2000) & 02:43:45.70 \\
Dec. (J2000) & -28:40:40.08 \\
Redshift $z$ & 0.0742 \\
$d_L$ & 347.0 Mpc \\\hline
$\log M_*/M_\odot$ & $10.28^{+0.06}_{-0.14}$ \\
$t_{\rm age}$ [Gyr] & $1.6^{+5.4}_{-0.4}$ \\
$\tau_2$ & $0.69^{+0.034}_{-0.028}$ \\
$[{\rm M/H}]$ & $-1.23^{+0.14}_{-0.16}$ \\
$t_{\rm burst}$ & $0.3^{+5.4}_{-0.2}$ \\
$f_{\rm burst}$ & $0.78^{+0.11}+{-0.34}$ \\
$\log M_{\rm BH}/M_\odot$ (from $M_{\rm BH}-M_*$) & $6.94 \pm 0.82$ \\
\enddata
\tablecomments{R.A. and Dec. are from the Legacy imaging survey \citep[][]{York2000}. Redshift is as measured in our work. The parameters below the line are derived from an SED fit using \texttt{fsps} and \texttt{prospector} \citep[][]{Johnson2021, Conroy2009, Conroy2010}. dust2 is the  SMBH mass is measured using the \cite{Greene2020} $M_{\rm BH}-M_*$ relation.  }
\end{deluxetable}

VT J0243 is offset by 0.2\arcsec ($1\sigma$ uncertainty $\sim0\farcs15$) from the Pan-STARRS centroid of the galaxy 2dFGRS TGS314Z138. 2dFGRS TGS314Z138 is an SA galaxy. We summarize relevant properties of this host in Table~\ref{tab:host}, including its redshift and location. In this section, we will constrain the star formation rate, stellar and black hole mass, and BPT classification of this galaxy. We will use these properties to constrain the origin of the emission associated with VT J0243 and the trigger of VT J0243 later in this work. 

To measure the stellar mass and star formation history of the galaxy, we performed an SED fit using the \texttt{Prospector} code \citep[][]{Johnson2021,Conroy2010, Conroy2009} and the WISE, GALEX, and Pan STARRS galaxy photometry following a similar procedure to \cite{Somalwar2021} and references therein. We assume a tau-model star formation history (SFR$\propto e^{-t/\tau}$), a \cite{Chabrier2003} IMF, and extinction following \cite{Calzetti2000}. We use \texttt{emcee} \citep[][]{Foreman-Mackey2013} to fit the SED, with 100 walkers, 500 burn-in steps, and 50000 steps. The results showed that $\tau$ was very small, with a posterior distribution rising towards $\tau < 0.1$ Gyr and flattening for lower values. \texttt{prospector} does not support such low values of $\tau$, so we reran the fit including a burst component (i.e., a delta function of star formation) and fixing $\tau = 0.1$ Gyr. We found the fraction of the stellar mass formed in the burst was poorly constrained but peaked towards $1$. The age of the burst is also poorly constrained. We report the maximum-a-posteriori estimate and $1\sigma$ highest posterior density interval for each fit parameter in Table~\ref{tab:host}

First, we consider the star formation rate of this galaxy. The star formation rate is critical for constraining the source of the observed X-ray emission (Section~\ref{sec:xray}). Our SED fitting results were consistent with a large fraction of the stellar mass forming in a star formation burst near the lookback time at $z=0.0742$. Hence, the star formation rate could be very high for this source ($\gtrsim 1\,M_\odot\,{\rm yr}^{-1}$). However, our constraints are sufficiently loose that the SFR may be $\ll 1\,M_\odot\,{\rm yr}^{-1}$. For galaxies with star formation that has remained constant for ${\sim}6$ Myrs, the H$\alpha-$SFR relationship can be used to set an upper limit on the SFR as SFR$=5.5\times10^{-42} L_{\rm H\alpha}\sim 0.21\,M_\odot\,{\rm yr}^{-1}$, with ${\sim}15\%$ uncertainty \cite{Calzetti2013}. We measured $L_{\rm H\alpha} = (3.73 \pm 0.05)\times 10^{40}$ erg s$^{-1}$, before any host extinction corrections. If we use the H$\alpha$/H$\beta$ ratio to measure the host extinction, we find $A_{\rm H\alpha} = (3.33 \pm 0.80) \times 1.97 \log \bigg( \frac{{\rm H\alpha}/{\rm H\beta}}{2.86}\bigg) = 1.09 \pm 0.26$ \citep[][]{Dominguez2013}. Then, the extinction corrected H$\alpha$ luminiosity is $L_{\rm H\alpha,0 } = (10.2 \pm 2.5)\times 10^{40}$ erg s$^{-1}$. Plugging this luminosity into the H$\alpha$-SFR relationship, we find SFR$=0.56 \pm 0.16$; in other words SFR$ <1\,M_\odot\,{\rm yr}^{-1}$ at the $3\sigma$ level. We find a similar constraint using the [O\,II] luminosity and the SFR-[O\,II] relation from \cite{Kewley2006}. These constraints are robust even if the line emission is not entirely produced by star formation (see the end of this section for a discussion of possible AGN activity in 2dFGRS TGS314Z138). However, the H$\alpha$ constraint relies on the assumption that the star formation has been constant for at least 6 Myrs.

Next we use the stellar mass of 2dFGRS TGS314Z138 to constrain the black hole mass of the galaxy. The black hole mass is critical for constraining the origin of VT J0243, as different types of transients dominate at different masses (e.g., TDEs cannot occur for $M_{\rm BH} \gtrsim 10^8\,M_\odot$). The stellar mass of this galaxy is well-constrained at $\log M_* = 10.06_{-0.08}^{+0.12}$. Using the black hole-stellar mass relation from \cite{Greene2020}, we find a black hole mass $\log M_{\rm BH}/M_\odot = 6.94 \pm 0.82$.

\begin{figure*}
\gridline{\fig{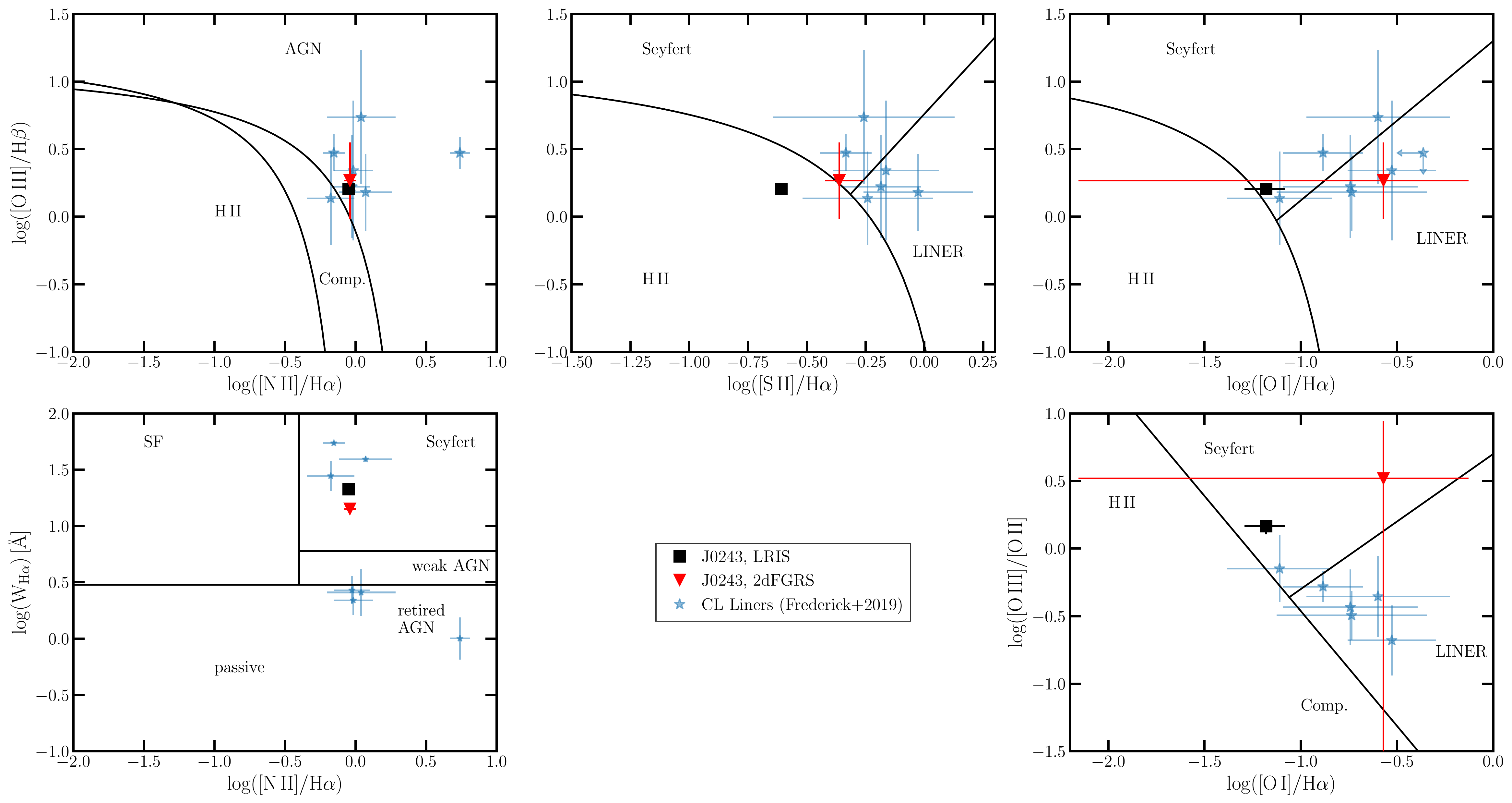}{0.95\textwidth}{}}
\caption{ Five versions of the BPT diagram \citep[][]{Baldwin1981, Kewley2006, Cid2011}, following Figure 13 from \cite{Frederick2019}. Line ratios measured from the LRIS (2dFGRS) observations of 2dFGRS TGS314Z138 are shown as black squares (red trangles). 2dFGRS TGS314Z138 is consistent with a Seyfert in most of the diagrams. \label{fig:BPT}}
\end{figure*}

Finally, we discuss the optical spectral features, and use them to classify 2dFGRS TGS314Z138 as a possible Seyfert galaxy. Both the archival 2dFGRS and the LRIS spectra show narrow line emission associated with AGN, such as $[{\rm O\,III}]\,\lambda\lambda 4959, 5007$, H$\beta$, H$\alpha$, $[{\rm N\,II}]\lambda\lambda 6548,6583$, and $[{\rm S\,II}]\lambda\lambda6716,6731$. We fit the spectra and measure emission line strengths using the same procedures as \cite{Somalwar2021}, and we refer the reader to that work for details. There has been no significant evolution in any of the line ratios, so we assume that the recent LRIS spectrum does not include any transient emission associated with VT J2043 and place both spectra on BPT diagrams \citep[][]{Baldwin1981, Kewley2006, Cid2011}, as shown in Figure~\ref{fig:BPT}. Both spectra are consistent with BPT-weak Seyferts. Likewise, none of the available WISE mid-infrared colors of 2dFGRS TGS314Z138 are consistent with a strong AGN \citep{Assef2018}. We thus identify this galaxy as a weak Type 2 Seyfert. As we will discuss in Section~\ref{sec:xray}, this galaxy may be a {\it true} Type 2 Seyfert, meaning that the absence of broad lines may be due to the complete lack of a broad line region (BLR). From the ROSAT soft X-ray flux constraints, we can constrain the pre-flare AGN accretion rate. Since $L_X \lesssim 2.6\times10^{42}$ erg s$^{-1} \lesssim 10^{-3} L_{\rm edd.}$ and assuming a bolometric correction ${\sim}20$ \citep[][]{Lusso2012}, we find $f_{\rm edd.} \lesssim 2\%$. 


\subsection{Radio analysis} \label{sec:radio}

\begin{figure*}
\gridline{\fig{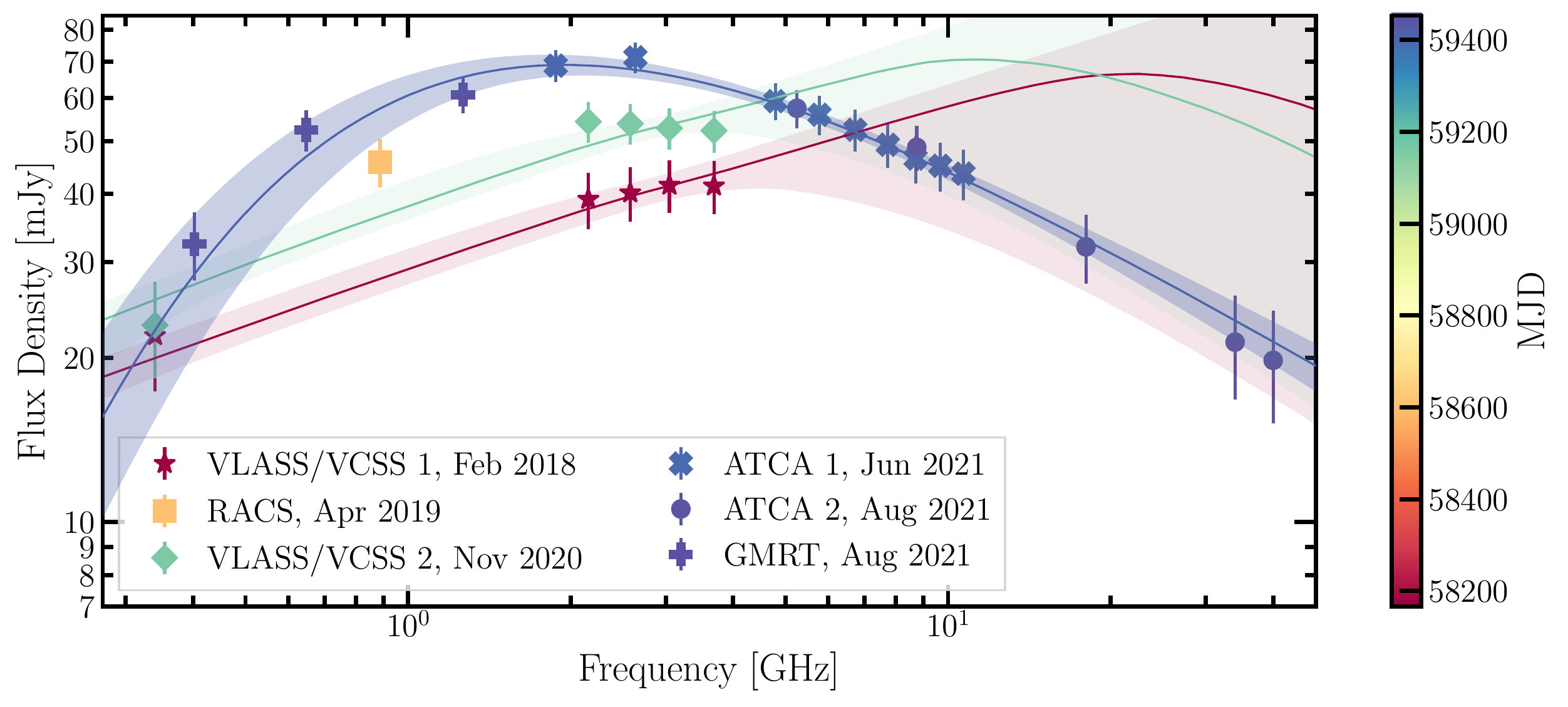}{0.95\textwidth}{}}
\caption{ The evolution of VT J0243's radio SED. The circles with errorbars correspond to our observations, and are colored with the observation epoch. We fit a synchrotron model to the three main observation epochs (see Section~\ref{sec:radio} for details), and show the best-fit model and $1\sigma$ errors as lines and bands. \label{fig:radio}}
\end{figure*}

Typically, radio emission from galactic centers is dominated by synchrotron emission due to particles accelerated within a relativistic, collimated jet \citep[][]{Blandford2019} or shocks from the collision of a jet and/or non-relativistic, wide-angle outflow with the circumnuclear medium (CNM). This emission can be self-absorbed or free-free absorbed. Because we are observing a transient, the outflow or jet must be expanding. In this section, we combine a fit to our VLBA observations of this source with synchrotron modelling of the observed SED (Figure~\ref{fig:radio}) to constrain the physical parameters of the source. 

\subsubsection{Synchrotron analysis methods}

 We constrain the physical properties of the source by assuming equipartition between the energy in electrons and the energy in the magnetic field. We also adopt the standard assumption that the relativistic electron distribution is a power-law in Lorentz factor (LF) above a minimum LF $\gamma_m$: $N(\gamma) d\gamma = N_0 \gamma^{-p} d\gamma,\,\gamma > \gamma_m$. In this case, the SED is well-modelled by a broken power law \citep[][]{Chevalier1998, Granot2002}. 
 
The slopes of the power law segments depend on the ordering of a number of characteristic LFs. The three relevant LFs for this work are (1) the LF of the lowest energy electrons $\gamma_m$, (2) the electron energy at which the optical depth to synchrotron self-absorption is one, $\gamma_{\rm sa}$, and (3) the energy at which the electron cooling timescale is shorter than the age of the source, $\gamma_c$. Each of these corresponds to a characteristic synchrotron frequency $\nu_{\rm x} = \gamma_{\rm x}^2 eB/(m_ec),\,{\rm x}\in[m,\,{\rm sa},c]$, where $e$ is the electron charge, $B$ is the magnetic field strength, and $m_e$ is the electron mass. These characteristic frequencies correspond to the locations of the breaks in a multiply-broken power law model of the synchrotron emission.

Our radio SED at all epochs is best-modelled when $\nu_{\rm sa} < \nu_m < \nu_c$. No other orderings can reproduce the observed broad and flat peak. For $\nu_{\rm sa} < \nu_m < \nu_c$, the power-law slope in the optically thick regime ($\nu < \nu_{\rm sa}$) is 2, corresponding to the slope of a Rayleigh-Jeans law with constant brightness temperature. For $\nu_{\rm sa} < \nu < \nu_m$, the slope is $1/3$, which is that of a single electron spectrum at frequencies smaller than the characteristic synchrotron frequency of that electron. For $\nu_m < \nu$, the power-law slope is $\alpha = -(p-1)/2$. Since each electron primarily emits at its characteristic synchrotron frequency $\nu \propto \gamma^2$ and the synchrotron power for a single electron $-\frac{dE}{dt} \propto \gamma^2$, we can approximate the flux density $S_\nu(\nu) d\nu = -\frac{dE}{dt} N(\gamma) d\gamma \propto \gamma^{2-p} d\gamma \propto \nu^{-(p-1)/2} d\nu$, leading to the slope $\alpha = -(p-1)/2$.

Following \cite{BarniolDuran2013}, we can now derive expressions for the number of electrons in the outflow $N_e$, magnetic field $B$, and total energy $E$ as a function of radius $R$, bulk Lorentz factor $\Gamma$, and radio SED properties. We assume a fraction $\epsilon_e$ of the total energy is stored in electrons, and a fraction $\epsilon_B$ is stored in the magnetic field. We nominally assume $\epsilon_e = \epsilon_B = 0.1$, although at the end of this section we will vary those values. We assume the outflow has an area $f_A \pi R^2/\Gamma^2$ and volume $f_V \pi R^3/\Gamma^4$. As we will discuss, VT J0243 has transitioned to a regime where $\Gamma \sim 1$, so the following equations apply to the nonrelativistic limit.
\begin{gather}
    N_e = \frac{9cF_p^3d_L^6\eta^{\frac{10}{3}}\Gamma^2}{2\sqrt{3}\pi^2e^2m_e^2\nu_p^5(1+z)^8f_A^2R^4} \approx 3.1\times10^{52} \frac{F_{p,-25}^3\eta^{\frac{10}{3}}\Gamma^2}{\nu_{p,9}^5R_{18}^4}, \\
    B = \frac{8\pi^3m_e^3c\nu_p^5(1+z)^7f_A^2R^4}{9eF_p^2d_L^4\eta^{\frac{10}{3}}\Gamma^3} \approx 0.18\,{\rm G}\bigg( \frac{\nu_{p,9}^5R_{18}^4}{F_{p,-25}^2\eta^{\frac{10}{3}}\Gamma^3}\bigg), \\
    E =  \frac{1}{\epsilon_B}\frac{f_V R^3 B^2}{8 \Gamma^2} \approx 5.3\times10^{51}\,{\rm erg}\bigg( \frac{1}{\epsilon_B}\frac{R_{18}^3 B_{-0.74}^2}{\Gamma^2}\bigg).
\end{gather}
Here, $d_L$ is the luminosity distance, $F_p$ is the peak flux density, $\nu_p$ is the peak frequency, and $z$ is redshift. The notation $Y_{x}$ denotes quantity $Y$ in units of $10^x$ cgs. The variable $\eta$ is defined as the ratio between the minimum and self-absorption frequencies: $\eta = \nu_m/\nu_{\rm sa}\,\, {\rm if}\,\,\nu_a < \nu_m;\,\,{\rm else}\,\,1$. Only the final equation for total energy $E$ assumes equipartition. In the final equalities, we have adopted the luminosity and redshift of VT J0243. We also assume, both in these equalities and henceforth, that $f_A=1$ and $f_V=4/3$, appropriate for a spherical, nonrelativistic outflow. For a jet, the appropriate values are $f_A = f_V = (\theta_j \Gamma)^2$, where $\theta_j$ is the jet half-opening angle. For the jetted TDE Sw J1644, assuming $\theta_j \sim 0.1$, we have $f_A,\, f_V \gtrsim 0.1$ \citep[][]{Eftekhari2018}. 

We will also require the electron density of the material into which the outflow is expanding. We denote this density $n_e$. Note that $n_e \neq N_e/(4\pi R^3)$, since $N_e$ is the electron number {\it in} the outflow, whereas $n_e$ is the density of the material {\it outside} the outflow. We derive this density following \cite{Ho2019}, who require conservation of momentum across the shock front and find 
\begin{equation*}
    n_e = \frac{B^2}{6\pi \epsilon_B \beta^2 c^2 m_p} \approx 1.1\,{\rm cm^{-3}}\bigg( \frac{\nu_{p,9}^{10}R_{18}^8}{F_{p,-25}^4\eta^{\frac{20}{3}}\Gamma^6 \beta^2 \epsilon_B}\bigg).
\end{equation*}
Here, $\beta = v/c$, where $v$ is the outflow velocity. This equation assumes expansion of a thermal gas ($\gamma = 5/3$) into fully ionized hydrogen. The generalization to a relativistic gas would modify this equation by a factor of order unity, which we neglect as it is significantly smaller than our measurement errors.

To apply these equations, we require a measurement of $\eta$. The self-absorption frequency of this system is likely at or below the low frequency end of our observations, so we cannot tightly constrain $\eta$ by fitting for $\nu_{\rm sa}$ and $\nu_{\rm m}$. Instead, we use the outflow size measured using our VLBA observations. Under equipartition, the outflow radius is related to $\eta$ as
\begin{gather*}
    R = \bigg[\frac{3^7c}{32\sqrt{3}\pi^9m_e^8}\frac{p-1}{p-2}\frac{\epsilon_B}{\epsilon_e}\bigg]^{\frac{1}{17}}\frac{\Gamma^{\frac{11}{17}}F_p^{\frac{8}{17}}\eta^{\frac{35}{51}}d_L^{\frac{16}{17}}}{f_V^{\frac{1}{17}}(1+z)^{\frac{25}{17}}f_A^{\frac{7}{17}}\nu_p} \\ \nonumber
    \approx 6.1\times10^{17}\,{\rm cm} \bigg[\bigg(\frac{p-1}{p-2}\frac{\epsilon_B}{\epsilon_e}\bigg)^{\frac{1}{17}}\frac{\Gamma^{\frac{11}{17}}F_{p,-25}^{\frac{8}{17}}\eta^{\frac{35}{51}}}{(1+z)^{\frac{25}{17}}\nu_{p,9}}\bigg].
\end{gather*}

\subsubsection{Synchrotron analysis results}

\begin{deluxetable*}{cc||c|ccccc}
\tablecaption{Synchrotron Analysis Results \label{tab:synchpars}}
\tablehead{\colhead{$p$} & \colhead{$\frac{R}{\rm pc}$} & \colhead{$\epsilon_e,\epsilon_B$} & \colhead{$\Gamma$} & \colhead{$\eta$} & \colhead{$\log \frac{B}{10^{-2} \rm G}$} & \colhead{$\log \frac{n_e}{{\rm cm}^{-3}}$} & \colhead{$\log \frac{E}{{\rm erg\,s}^{-1}}$} }
\startdata\hline
\multirow{4}{*}{$2.2\pm0.12$} & \multirow{4}{*}{$0.71\pm0.02$} & \multirow{2}{*}{$0.1,0.1$} & 1.005 & $5.2\pm1.1$  & $2.74\pm0.82$ & $1.32\pm0.16$ & $52.02\pm0.16$ \\
 & & & 1.3 & $4.05\pm0.89$ & $3.0\pm0.2$ & $-0.11\pm0.20$ & $51.88\pm0.20$ \\\cline{3-8}
 & & \multirow{2}{*}{$0.1,10^{-3}$} & 1.005 & $5.2\pm1.1$ & $0.70\pm0.15$ & $2.23\pm0.14$ & $52.81\pm0.14$ \\
 & & & 1.3 & $4.05\pm0.89$ & $0.75\pm0.23$ & $0.69\pm0.17$ & $52.64\pm0.17$
\enddata
\centering 
\end{deluxetable*}

In this section, we present the physical parameters derived from our synchrotron analysis. First, however, we consider the fact that our observations can only be fit in the regime  $\nu_{\rm sa} < \nu_m < \nu_c$. This is unusual $-$ the \cite{Granot2002} model for an adiabatically expanding outflow applied to this source suggests that we should observe $\nu_{\rm sa} > \nu_m$ given the $>1000$ day age of the outflow. A very high $\nu_m$ at late times requires a source of energy which keeps the electron population at high $\gamma$. Thus, continual energy injection could explain our observation of $\nu_{\rm sa} < \nu_m$. Continual energy injection is also a possible explanation of the unusual, rising late-time radio light curve (Figure~\ref{fig:lc}; also see Section~\ref{sec:disc})

To derive the physical parameters, we first constrain $F_p$ and $\nu_p$. We fit a doubly-broken power-law to the most recent observation epochs (GMRT+ATCA 1+ATCA 2) using the \texttt{dynesty} software \citep[][]{Speagle2020}. We fix the slopes to the expected values described above, and allow the position of each break and the electron spectral index $p$ to float. We adopt broad, Heaviside priors on all parameters, except for $p$, which we require be in the physically-motivated range $[2,5]$. We use the resulting best-fit model to evaluate the peak flux density and frequency, along with their uncertainties. We find $F_p = 66.7 \pm 3.1$ mJy, $\nu_p = 2.28 \pm 0.28$ GHz, and $p = 2.40 \pm 0.17$. Note that the peak frequency is consistent within $<2.5\sigma$ with the best-fit characteristic minimum frequency, $\nu_m = 2.78 \pm 0.35$ GHz. Next, we constrain the bulk lorentz factor, $\Gamma$. The outflow was launched before the first VLASS observation epoch on MJD 58166; hence, it is at least ${\sim}1400$ days old. It was launched after the NVSS observation, so it is no more than ${\sim}10000$ days old, Thus, we have $0.1 < \langle \beta \rangle \lesssim 0.6$ and average bulk Lorentz factor $1.005 < \Gamma \lesssim 1.3$. 

Next, we calculate $\eta$ using the outflow size and the equipartition radius equation. As described in Section~\ref{sec:obs}, our VLBA observations imply an approximate radius $R = 0.71 \pm 0.02$ pc. Thus, we have $\eta = 5.17 \pm 1.13\,(\Gamma = 1.005), 4.05\pm 0.89\,(\Gamma=1.3)$. In both cases, the predicted $\nu_{\rm sa}$ is consistent with constraints from our doubly-broken power-law fit. We reran the doubly-broken power-law fit while requiring $\eta$ be consistent with the above values and found that $p$ has not changed significantly from our previous measurement: $p=2.20\pm 0.12$.

Finally, we constrain the magnetic field, electron number density, and total energy. The results are tabulated in Table~\ref{tab:synchpars}. The measured densities are consistent with results for other galaxies: at a similar distance (in units of the Schwarzschild radius), typical densities are $\gtrsim 10^{-1}$ cm$^{-3}$ (see Fig. 2 of \citealp{Alexander2020}). The energies are consistent with jetted TDE observations \citep{Eftekhari2018}.

Our assumption of $\epsilon_B = 0.1$ has been shown to be incorrect for the jetted TDE Sw J1644+57 \citep{Eftekhari2018}. If we adopt the preferred value for that event, $\epsilon_B = 10^{-3}$, our physical parameters are modified, and the results are listed in Table~\ref{tab:synchpars}. The energy is now higher than measured for previous jetted TDEs \citep{Eftekhari2018}. A collimated geometries (i.e., smaller $f_A$ and $f_V$) will tend to decrease the energy ($E\propto f_V^{3/7}$), increase the magnetic field ($B\propto f_V^{3/7}$), and increase $n_e$ ($n_e \propto f_V^{6/7}$). 

The evolution of VT J0243's radio SED is shown in Figure~\ref{fig:radio}. The datapoints are colored by the observation MJD. We have overplotted a doubly-broken power law fit to the most recent epoch in purple. We overplot fits to the VLASS/VCSS observations. In these fits, $p$ is forced to be consistent with the value measured from the most recent observations. The break frequencies and amplitude are allowed to float freely. These observations are not sufficiently well-sampled to provide strong constraints on any physical parameters, but are roughly consistent with expectations for an expanding outflow.

In summary, VT J0243 is associated with a luminous, energetic outflow. The outflow is currently non-relativistic, but given the high, and still rising, luminosity, we believe it likely that we are observing a relativistic jet, possibly off-axis, that has slowed. This hypothesis is supported by the observed non-spherical geometry from the VLBA (see the end of Section~\ref{sec:radioobs}).

\subsection{X-ray analysis} \label{sec:xray}

In this section, we discuss our X-ray observations. First, we present constraints on X-ray emission at the time that VT J0243 turned on. Then, we discuss the luminosity and spectrum from our more recent X-ray observations. Finally, we consider three possible origins for this late-time X-ray emission: star formation, an accretion disk, or something associated with the transient event.

Beginning on MJD ${\sim}59375$, we detected near-constant X-ray emission from the location of VT J0243 with a $0.3-10$ keV luminosity of $\log L_{\rm 0.3-10\,keV}/({\rm erg\,s}^{-1}) = 42.3 \pm 0.01$ (Figure~\ref{fig:lc}), after correcting for Milky Way H\,I absorption ($N_{H, {\rm MW}} = 1.51\times10^{20}$ at the location of VT J0243; \citealp{HI4PI2016}). We do not have strong constraints on the X-ray emission before that date, although from the archival MAXI observations (black triangle upper limits) we can rule out X-ray emission with the same luminosity and lightcurve of the jetted TDE Sw J1644 (red line). We cannot rule out a flare with average luminosity over ${\sim}100$ days that is ${\lesssim}2L_{\rm edd.}$.

\begin{figure}
\gridline{\fig{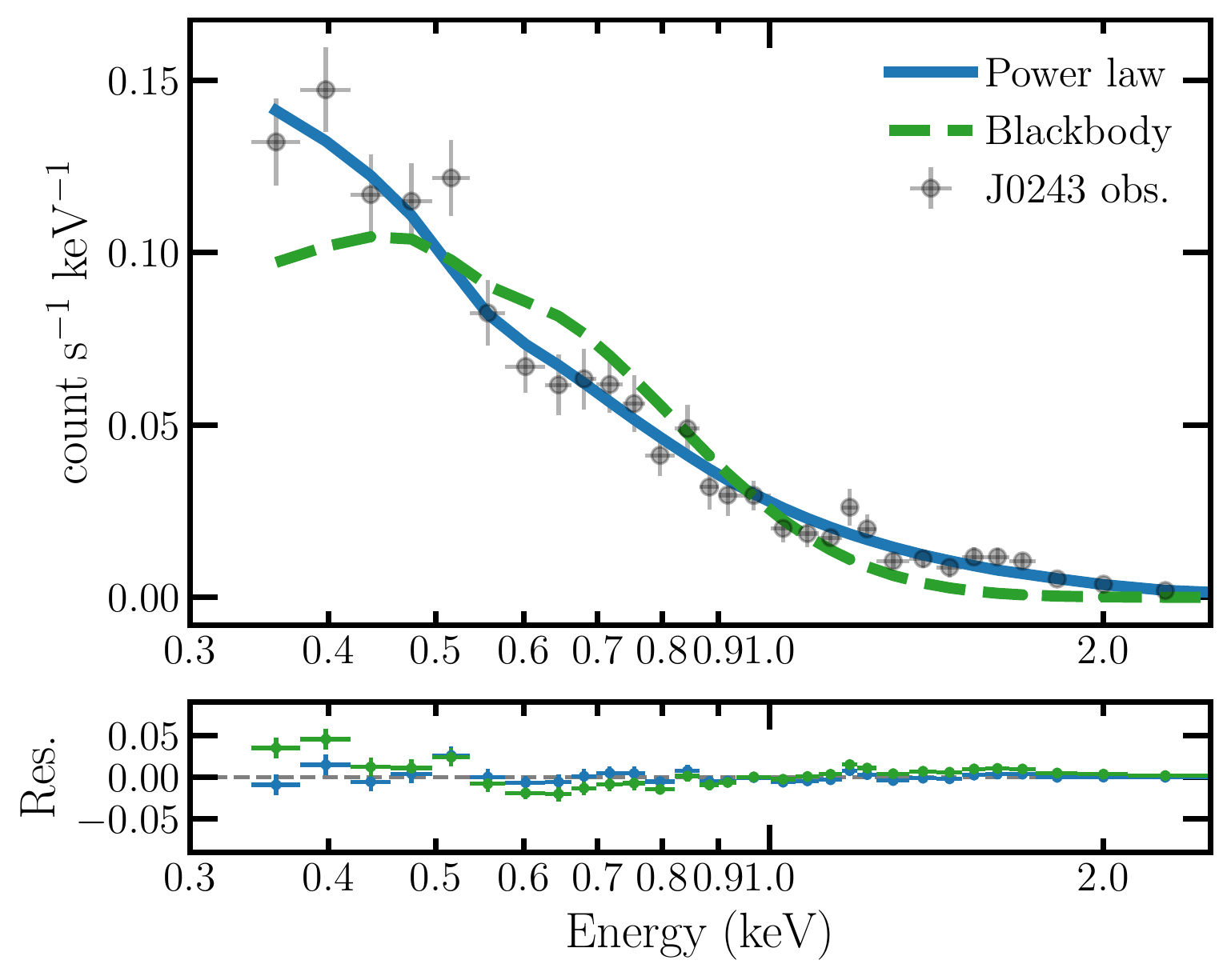}{0.45\textwidth}{}}
\caption{ The {\it XMM}-Newton x-ray spectrum for 2dFGRS TGS314Z138. The observations are the black points, while the lines show model fits. The emission is most consistent with a power law model with little intrinsic hydrogen column density. \label{fig:xray_spec}}
\end{figure}

The late-time X-ray spectrum is shown in Figure~\ref{fig:xray_spec}. We used \texttt{xspec} to fit the X-ray emission to an absorbed power law (\texttt{cflux*TBabs*zTBabs*powerlaw}) and a blackbody (\texttt{cflux*TBabs*zTBabs*bbody}).
In both cases, we include both Milky Way extinction, for which we fix the hydrogen  column density to the known value $N_{\rm H, MW} = 1.52 \times 10^{20}$ cm$^{-2}$ \citep[][]{HI4PI2016}, and intrinsic extinction, for which we let the Hydrogen column density float. The best-fit models are shown in Figure~\ref{fig:xray_spec}. The pure blackbody cannot fit our observations (\texttt{cstat}/$dof=195/32$), but 
power law (\texttt{cstat}/$dof=33.7/33$) provides a statistically acceptable fit. 
The best-fit power law parameters are: intrinsic column density $<7.7\times10^{19}$ cm$^{-2}$ ($5\sigma$), photon-index $\Gamma = 2.98 \pm 0.06$, and an absorbed $0.3-10$ keV flux density $\log f_{\rm 0.3-10\,keV} = -12.86 \pm 0.01$ ($L_{\rm 0.3-10\,keV}/({\rm erg\,s}^{-1}) = 42.3 \pm 0.01$). We will discuss the interpretation of these parameters later in this section.

We consider three general categories of X-ray sources: (1) star formation in the host galaxy, (2) an accretion disk with or without a hot electron corona, (3) other transient emission associated with VT J2043. We will now discuss the likely contribution of each of these sources in turn.

X-ray photons associated with star formation are predominantly emitted by low- and high-mass X-ray binaries (LMXBs/HMXBs; \citealp{Mineo2014}). The star formation rate is correlated with the $2-10$ keV X-ray luminosity as SFR$=(1.40\pm 0.32)\times \frac{L_{\rm 2-10\,{\rm keV}}}{10^{40}\,{\rm erg\,s}^{-1}}\,M_\odot$ yr$^{-1}$ \citep{Vattakunnel2012}. To reproduce the observed $2-10$ keV luminosity of $4.7\times10^{40}$ erg s$^{-1}$, we require that the SFR$=6.58 \pm 1.5\,M_\odot$ yr$^{-1}$. This is consistent with our SED fit, but our SED fit provides very weak constraints on the SFR. It is also consistent with our pre-flare radio limits: radio emission due to star formation has been empirically measured to be SFR$ = 5.52\times 10^{-22} L_{\rm 1.4\,GHz,\,SFR}$ for $L_{\rm 1.4\,GHz,\,SFR} > 6.4\times 10^{28}$ erg s$^{-1}$ Hz$^{-1}$, which corresponds to $L_{\rm 1.4\,GHz,\,SFR} = 0.83 \pm 0.22$ mJy for SFR$=6.58 \pm 1.5$. This star formation rate is inconsistent with the observed H$\alpha$ emission: from Section~\ref{sec:host}, the SFR based on the H$\alpha$ emission is SFR$=0.56\pm0.16$. This expression for the SFR-H$\alpha$ correlation is only valid if the SFR has been ${\sim}$ constant for $>6$ Myr, but, if the star formation was very recent, we would not expect to see the X-ray emitting LMXBs and HMXBs. Hence, it is unlikely the X-ray emission was produced by star formation. We briefly consider alternative X-ray sources in the rest of this section.  

X-ray emission from AGN is dominated by thermal emission from the disk, and inverse Comptonized thermal photons by the hot electron corona \citep[][]{Haardt1991}. Using the observed correlation between narrow [O\,III]$\lambda$5007 and H$\alpha$ with X-ray luminosity \citep{Netzer2006}, an accretion disk can account for all of the observed X-ray flux. Given that the X-ray lightcurve is consistent with a constant luminosity, it is feasible that the X-ray emission is entirely due to an active accretion disk. A $\Gamma \sim 3$ power law spectrum is consistent with observations of narrow line Seyfert 1 galaxies \citep[][]{Grupe2010}, although radio-loud Seyferts typically have flatter X-ray spectra ($\Gamma \sim 2$; \citealp{Komossa2018}), but with large scatter. 

The X-ray emission from 2dFGRS TGS314Z138 is not entirely consistent with ``normal'' Seyferts. The intrinsic column density is consistent with zero and inconsistent with the $n_{\rm H} > 10^{22}$ cm$^{-2}$ typically measured in Seyfert 2s \citep[][]{Risaliti1999}. For standard AGN, such a low gas column density means that the BLR should be observable \citep[][]{Panessa2002}. However, as discussed in Section~\ref{sec:host}, we do not detect any broad line emission. This low $n_{\rm H}$ may support the hypothesis that the X-ray emission is dominated by star formation. If it is not, and the column density is truly near-zero, 2dFGRS TGS314Z138 may be a ``true'' Seyfert 2, which show small X-ray column densities but no broad line emission \citep[][]{Hawkins2004}. 
Finally, we consider the scenario where the observed X-rays are transient, rather than associated with an old accretion disk or star formation, and consider a few of the possible origins. X-rays may be emitted from the forward shock of the outflow, as may have been the case for jetted TDEs like Sw J1644+57 \citep{Eftekhari2018}. In this case, we expect the X-ray slope to be $\Gamma = p/2+1 = 2.1 \pm 0.05$, where $p$ is taken from Table~\ref{tab:synchpars}. This $\Gamma$ is significantly inconsistent with our measured value. bremmstrahlung associated with the electrons in the radio-emitting outflow and dense clumps of CNM gas could produce X-rays, but we would expect a harder power-law spectrum in this case.

The X-rays may be associated with a new corona and associated accretion disk, formed as a result of, e.g., a stellar disruption. The observed power-law spectrum would be consistent with expectations for a transient corona/disk \citep[][]{Osterbrock1991}.

In summary, VT J0243 is not associated with an extraordinarily bright X-ray flare as has been observed for the extremeley luminous, on-axis, jetted TDEs. We cannot rule out a flare with $L\lesssim 2L_{\rm edd.}$. VT J0243 is detected in late-time X-ray observations with a $0.3{-}10$ keV luminosity $L_X\sim10^{42.3}$ erg s$^{-1}$, photon index $\Gamma \sim 3$, and negligible intrinsic column density. This emission is unlikely to be related to star formation. Instead, it is most likely a transient, or a pre-existing accretion disk.

\subsection{Infrared and optical analysis} \label{sec:OIR}

In Figure~\ref{fig:lc}, we show the infrared (bottom) and optical (middle) light curves for VT J0243. In this section, we will provide brief analyses of the possible origins of the observed transient emission. Because of the low cadence and insufficient sensitivity of the observations, we will not perform any detailed modelling.

There is a significant flare detected by the PanSTARRS survey near MJD${\sim}55895$ with $g$-band luminosity $L_g\sim4\times10^{42}$ erg s$^{-1}$. The flare brightened and faded over a timescale smaller than the PanSTARRS cadence ($\lesssim 400$ days). Given the low cadence, we cannot measure light curve shape in more detail, but it is consistent with optically-detected TDEs, which typically rise over tens of days and fade over ${\sim}60$ days \citep{vanVelzen2021}. The co-temporal CRTS observations detect the flare at a ${\sim}2\sigma$ level but are not sensitive enough to reliably constrain the lightcurve. They do suggest that the flare peaked around the time of the brightest PanSTARRS observation, so the peak luminosity is likely within a factor of a few of $L_g\sim4\times10^{42}$ erg s$^{-1}$. This is slightly dimmer than but consistent with typical optically-detected TDEs; the dimmest TDE from \cite{vanVelzen2021} peaked at $L_g \sim 7\times 10^{42}$ erg s$^{-1}$. We fit the fluxes to a blackbody assuming no intrinsic extinction, which is reasonable given the low column density measured from the X-ray spectrum (Section~\ref{sec:xray}). 

The optical fluxes at peak are consistent with a blackbody with no extinction and temperature $\log T_{\rm bb}/{\rm K} = 4.35 \pm 0.38$ and radius $\log R_{\rm bb}/{\rm pc} = -3.89 \pm 0.34$, corresponding to a blackbody luminosity $\log L_{\rm bb}/({\rm erg s}^{-1}) = 43.4 \pm 1.7$. Again, these blackbody parameters are all standard for optically-detected TDEs. We cannot rule out that the flare has repeated. Shortly after the peak, the optical emission rebrightens slightly to $L_g \sim (6.7\pm1)\times10^{42}$ erg s$^{-1}$.  The fluxes at the rebrightening are consistent with a blackbody with no extinction and temperature $\log T_{\rm bb}/{\rm K} = 3.70 \pm 0.07$ and radius $\log R_{\rm bb}/{\rm pc} = -3.19 \pm 0.20$, corresponding to a blackbody luminosity $\log L_{\rm bb}/({\rm erg s}^{-1}) = 42.3 \pm 0.5$. This blackbody luminosity is roughly an order of magnitude fainter than the brighter peak. The temperature is significantly cooler than the first peak, and the emission may come from a larger radius.

Unfortunately, because of the limited cadence of the PanSTARRS observations and the large uncertainties even at the optical peak, we cannot perform detailed modelling to determine the origin of the optical flare. The peak emission can be modelled as a standard accretion disk. It also could be a thermally-emitting outflow, heated by EUV emission from, e.g., an accretion disk, as may be observed in TDEs. 

The infrared lightcurve shows variability with an approximate amplitude ${\sim}150\,\mu$Jy (${\sim}10^{42}$ erg s$^{-1}$). The emission appears to redden slightly between the first three epochs and the rest of the MIR observations, which suggests that the MIR-emitting dust was heated when, e.g., the accretion rate increased. Intriguingly, the time period when this change must have occurred is roughly consistent with the range of launch dates constrained by the outflow radius evolution, and the time of the optical flare. The average change in flux density in each band between the first three epochs and the later observations is $\Delta f_{\nu}({\rm W1}) = 44.8 \pm 11.1\,{\rm \mu Jy},\,\Delta f_{\nu}({\rm W2}) = 88.5 \pm 17.6\,{\rm \mu Jy}$, where the uncertainties are determined through the standard deviation of the observations. 

If we assume that the dust started out cold and the entire flux change was due to the dust heating, we can fit the $\Delta f_{\nu}$ values during each WISE epoch to a blackbody to estimate the dust temperature and luminosity, albeit with large uncertainties and covariances. The average temperature over all WISE epochs is  $865\pm259$ K and the bolometric luminosity $\log L_{\rm IR}/({\rm erg\,s}^{-1}) \sim 41.69 \pm 0.15$. If we assume dust with a covering fraction ${\sim}1$, this implies it is located at an unrealistically small radius ${\sim}10^{-2}$ pc. Instead, we favor a scenario where more distant dust with a low covering fraction is located farther away (e.g., a covering fraction ${\sim}1\%$ corresponds to a distance ${\sim}0.1$ pc). Low covering fractions of ${\sim}1\%$ are consistent with measurements from IR flares during TDEs in quiescent galaxies \citep[][]{Jiang2021_IRTDE}. AGN typically have high covering fractions ${\gtrsim}40\%$ due to the dusty torus \citep[e.g.][]{Ricci2017}. The best fit average dust luminosity suggests that the bolometric luminosity of the EUV flare that heated the dust was $\log L_{\rm EUV}/{\rm erg\,s}^{-1} \sim 43.2 - \log (\frac{f_{\rm cov, dust}}{1\%})$. Unless the dust covering fraction is abnormally small, the EUV flare was sub-Eddington ($\lesssim 1\% L_{\rm edd.}$). 

This analysis of the IR flare assumes that there was no emission in between the low cadence WISE observations. A higher Eddington ratio EUV flare could have heated the dust between the IR observations, and we would not observe it. Hence, these constraints should be taken with a large grain of salt.

In conclusion, the low amplitude of the WISE variability suggests that either this galaxy has an extraordinarily low dust covering fraction, even when compared to completely quiescent galaxies, or that any EUV flare in the time period under question was sub-Eddington. There may have been a higher luminosity EUV flare in between the WISE observations. Moreover, this analysis has been subject to many poorly supported-assumptions. For example, if there was pre-existing accretion disk, our assumption that the dust was initially cold would be incorrect.

\begin{figure*}
\gridline{\fig{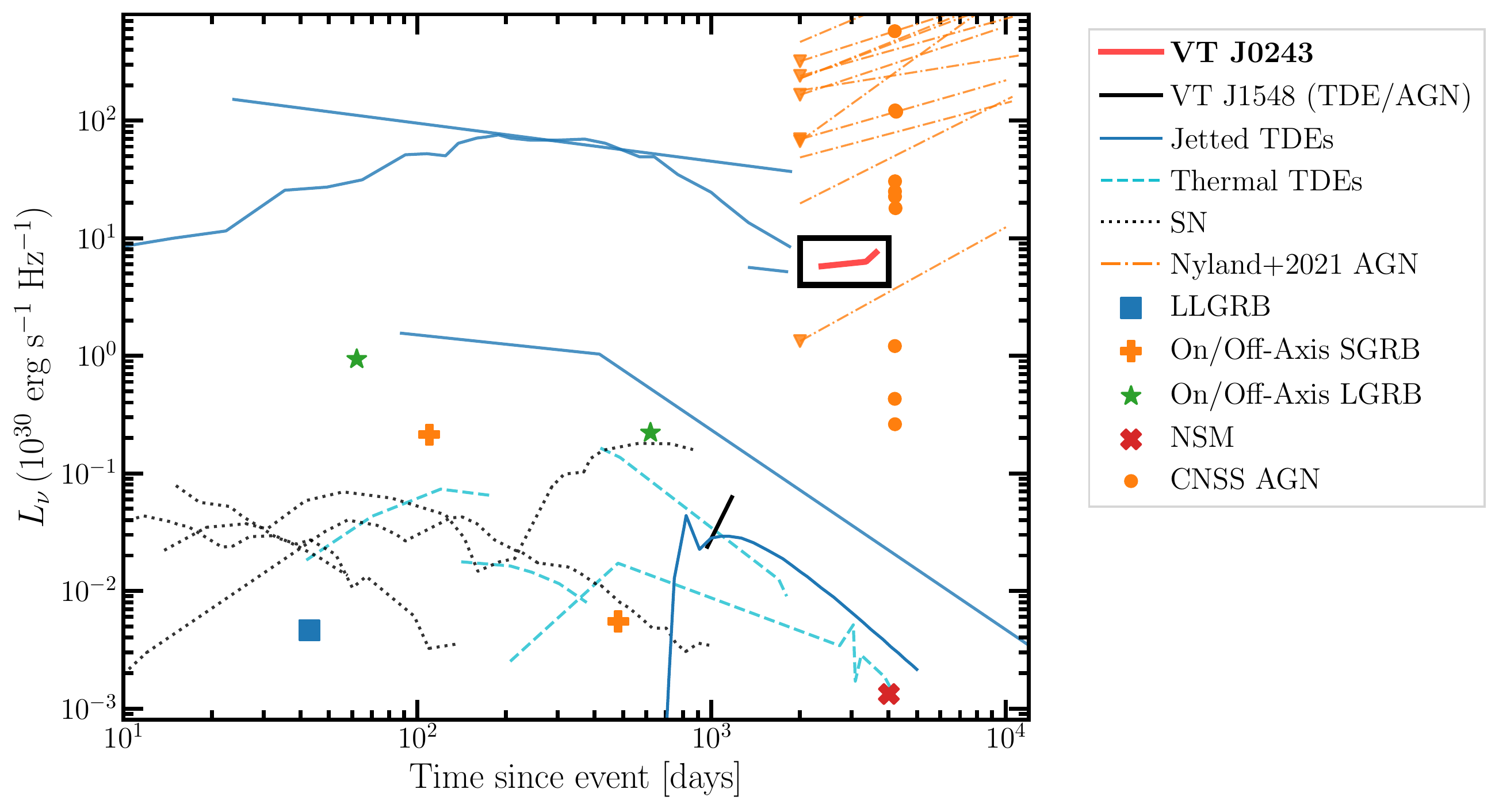}{1\textwidth}{}}
\caption{ Radio light curves (lines) and typical durations/luminosities (squares) for classes of radio transients. The squares are retrieved from \cite{Metzger2015} and are measured at 1.4 GHz. The TDE lightcurves are from \cite{Alexander2020, Ravi2021, Mattila2018}, and references therein, and are largely at 5 GHz. The SN lightcurves are at $5-8$ GHz, are from \cite{Salas2013, Soderberg2006, Soderberg2010, Kulkarni1998}. The lightcurve for VT J1548 is measured at 3 GHz \cite{Somalwar2021}. For comparison, the 5 GHz lightcurve of VT J0243 is shown in red, assuming the flare was launched around MJD 56000. This source is of comparable brightness to jetted TDEs, but is still rising whereas previous events began fading by ${\sim}1000$ days post-event. It is also at a comparable luminosity to the newly radio loud AGN from \cite{Wolowska2021}, shown as orange circles, and the radio variable AGN from \cite{Nyland2020}, which are shown as orange dot-dashed light curves. The triangles in the light curves denote upper limits. We have arbitrarily chosen the start date of these AGN flares for ease of comparison to VT J0243. \label{fig:lum_sum}}
\end{figure*}

\section{Discussion} \label{sec:disc}

The radio emission associated with VT J0243 is likely caused by the launching of a jet. Sub-relativistic outflows never produce the observed high luminosity radio emission ($\nu L_\nu ({\rm 3\,GHz}) > 10^{40}$ erg s$^{-1}$), nor such high energies ($E\sim10^{51}$ erg). Even fast ($\beta>0.1$), wide angle outflows from AGN are generally associated with radio-quiet sources and are compact ($\lesssim 0.1$ pc). Such outflows are not expected to be produced by disks with very low accretion rates, and no outflow has ever reproduced the observed outflow velocity, radio luminosity, and radio/X-ray luminosity ratio. Hence, we do not consider wide-angle, non- or semi-relativistic outflows further. Instead, we assume VT J0243 is associated with the launching of a jet. In this section we discuss the answer to the question: why did a jet launch? First, we summarize our observations:
\begin{itemize}
    \item X-ray emission with $\log L_X/{\rm erg\,s}^{-1} = 42.3\pm0.01$ and a power-law spectrum with index $\Gamma=2.98\pm0.06$ and a column density consistent with the Milky Way value. The emission is likely associated with a hot accretion disk and electron corona. The emission may have been transient or persistent, and may have evolved over the last few decades. The peak luminosity was likely $\lesssim 2 L_{\rm edd.}$.
    \item Transient radio emission with a current luminosity $\nu L_{\nu}({\rm 5\,GHz})=3.6\times10^{40}$ erg s$^{-1}$. The radio-emitting outflow is currently at a radius $R = 0.71 \pm 0.02$ pc. It has an average velocity $0.1 < \beta < 0.6$, or $1.005 < \Gamma < 1.3$, and is currently non-relativistic. It has a high equipartition energy ${\sim}10^{52}$ erg and a moderate electron density, ${\gtrsim}1$ cm$^{-3}$, depending on the assumed fraction of the energy stored in the magnetic field. The magnetic field is ${\sim}10^{-2}$ G.
    \item Significant optical variability, peaking at $L_g = 4\times10^{42}$ erg s$^{-1}$. The flare peaks around MJD 56000, and fades over ${\lesssim}400$ days. The peak is consistent with a $T\sim2\times10^4$ K blackbody with bolometric luminosity $10^{43.4}$ erg s$^{-1}$. After the flare faded, it rebrightened to a cooler blackbody ($T \sim 5000$ K) with a bolometric luminosity an order of magnitude dimmer at $10^{42.3}$ erg s$^{-1}$. 
    \item Weak MIR variability suggesting that any EUV flare in the last ${\sim}5000$ days was sub-Eddington, unless it occured between the low cadence IR observations. The MIR variability increased around MJD 56000, which is consistent with the launch date of the radio-emitting outflow if the outflow has travelled at a constant $\beta$.
    \item The host galaxy of VT J0243, 2dFGRS TGS314Z138, shows narrow line emission consistent with historic, weak Seyfert activity. The MIR colors, on the other hand, are consistent with quiescent galaxies.
\end{itemize}

In the rest of this section, we consider the possibility that this source is caused by a nascent jet associated with an accreting black hole. First, for completeness, we briefly discuss, and rule out, an alternate possibility: a supernova-triggered jet.

\subsection{Supernova-triggered jet}

Supernovae (SNe) can produce radio emission spanning from $L_\nu \sim 10^{25-32}$ erg s$^{-1}$ Hz$^{-1}$ for timescales as long as ten years \citep[][]{Weiler2002, Mooley2016}. The emission is often synchrotron emission associated with an outflow/jet colliding with the dense, local environment or a relativistic jet \citep[e.g.][]{Chevalier1998, Soderberg2010}. Typical SNe do not remain as bright as VT J0243 for such long periods of time (see Figure~\ref{fig:lum_sum}). Moreover, the $\mathcal{O}({\rm pc})$ size of the radio emitting outflow/jet associated with VT J0243 would be highly unusual. A gamma-ray burst (GRB) can produce such an outflow; however, no GRB has been observed with a rising radio luminosity thousands of days post-explosion \citep[e.g.][]{Kangas2021}. Moreover, the observation of $\nu_{\rm sa} < \nu_m$ thousands of days post-explosion is inconsistent with models of GRB outflow evolution \citep[][]{Granot2002}. Hence, VT J0243 is unlikely to be related to a supernova.

\subsection{Black hole accretion-triggered jet}

Accreting black holes, whether stellar mass or supermassive, are well established to be associated with jet activity. The process through which the jet is launched, the connection between the accretion disk and the jet, and the connection between the black hole properties (i.e., spin) and the jet remain open questions. In the following sections, we provide a basic summary of the physics of jets associated with black holes and accretion, and then we compare the properties of VT J0243 to those expected for young jets launched from accreting black holes. 

While the stellar mass black hole regime is not relevant to VT J0243, our understanding of jet physics and the disk-jet connection for stellar mass black holes is more sophisticated. We are better able to study these events because of the short timescales associated with the disk and jet evolution, which allow real-time observations of the jet and disk life cycles, and the smaller dynamic ranges of the systems, which allow for more realistic simulations. Ideally, the accretion disk and black hole evolution would be scale-free, so we can apply the same physics to stellar mass black holes and SMBHs. In reality, effects such as the mass-dependence of the inner disk temperature introduce a scale-dependence \citep[e.g.][]{Fender2007}. These effects have critical effects on accreting SMBHs, causing them to behave very differently in certain regimes (e.g., at very high accretion rates) from XRBs. Despite this, much of the stellar mass black hole physics is relevant to SMBHs, so we begin with a summary of stellar mass black hole disc/jet evolution. Then, we discuss the SMBH regime, and finally focus the discussion to comparisons with VT J0243.

\subsubsection{X-ray binary disk-jet connection}

The evolution of accreting stellar-mass black hole systems, X-ray binaries (XRB), is best understood by considering the evolution in X-ray hardness/luminosity space. When the X-ray binary is extremely sub-Eddington ($L_X/L_{\rm edd.} < 0.01$), the X-ray emission is low, with a flat spectral slope. Hence, this is called the low-hard state. In this low Eddington ratio regime, the accretion disk is geometrically-thick, optically-thin, and hot. It is radiatively inefficient, so advection dominates and this type of disk is called an advection dominated accretion flow (ADAF; \citealp{Narayan1994}). In the low-hard state, the XRB is typically observed to have a mildly relativistic ($\Gamma < 2$) jet \citep[][]{Fender2004}. 

As the Eddington ratio increases, the X-ray luminosity increases but the spectrum remains hard as the ADAF continues to dominate the disk. The radio luminosity likewise increases. Eventually, the X-ray emission reaches a peak, as the high Eddington ratio has caused the geometrically-thin outer-disk to extend into the inner disk and replace the ADAF. The X-ray spectrum softens, while the luminosity remains roughly constant \citep[][]{Fender2010}. During this softening, the jet Lorentz factor increases to $\Gamma > 2$, and the jet emission becomes intermittent and dominated by discrete blobs \citep[][]{Fender2010}. Soon after this change in the jet properties, the XRB will pass the ``jet line'', which is a characteristic hardness ratio at which the steady jet completely vanishes \citep[][]{Fender2009}. The XRB is now in the high-soft state. After this stage, the Eddington ratio will drop while the spectrum remains soft. At low Eddington ratios, the ADAF will begin to dominate again and the X-ray hardens. The XRB will cross the jet line again, and a new jet will launch. 

The processes through which the jet is quenched and launched are not fully understood. Both likely involve changes in the magnetic field in the accretion disk. The jet is likely collimated by pressure from external material; hence, the prevalence of jets in low Eddington ratio AGN with puffy disks \citep[][]{Tchekhovskoy2010}. The internal jet magnetic fields are generally unable to collimate more than the extreme base of the jet \citep[][]{Tchekhovskoy2009}. As we will discuss in Section~\ref{sec:tde}, jets are also sustainable near SMBHs accreting at near- or super-Eddington rates, as the disk again becomes puffy and the jet can be collimated. 

XRBs largely remain in the quiescent low-hard state, only entering the high-soft state during outbursts that are thought to be triggered by instabilities in the accretion disk \citep[][]{Fender2010}. There is some evidence that black hole spin is positively correlated with jet power, as would be expected if jets are powered by the \cite{Blandford1977} mechanism. However, the sample of XRBs with known spins remains small \citep[][]{Fender2010MNRAS}.

\subsubsection{The disk-jet connection for supermassive black holes}

There is observational evidence that the disk-jet connection for XRBs can be extrapolated to accreting SMBHs. For example, there is a tight, black-hole-mass dependent correlation between the X-ray and radio luminosities of XRBs, and observations of AGN have shown that these SMBHs lie on the same correlation \citep[][]{Gultekin2019}. Moreover, a modified version of the X-ray hardness-luminosity diagram, which replaces the X-ray hardness with the relative luminosity in power law and disk blackbody components, shows the same structure for XRBs and AGN \citep[][]{Fender2010}. It is not clear, however, that AGN follow the same cycle as XRBs in this diagram. The disk instabilities that cause XRB outbursts have not been proven to occur in AGN \citep[][]{Janiuk2011}. The relationship between spin and jet power is observationally unclear, as for X-ray binaries. The observed dichotomy between the radio-loud and quiet low-luminosity AGN (LLAGN) populations (${\sim}10\%$ of LLAGN are radio loud) is plausibly explained if the radio-quiet LLAGN have low SMBH spins while the radio-loud sources have extremal spins \citep[][]{Tchekhovskoy2010}. AGN simulations unambiguously find a strong, positive correlation between jet power and spin \citep[][]{Tchekhovskoy2010}.

As with XRBs, AGN with lower Eddington ratios ($\ll 0.1$) often have weak jets \citep[][]{Falcke2001, Fabian2012, Laha2021}. As we will discuss in Section~\ref{sec:tde}, there is strong evidence that accreting black holes at near- or super-Eddington rates also launch jets. For example, the TDE Sw J1644 launched a powerful jet during a period of near- or super-Eddington accretion. The exact mechanism through which this jet was launched is unconfirmed, but the observation of a jet from such a young accreting system suggests that the accretion disk became strongly magnetized remarkably quickly \citep[][]{Tchekhovskoy2014}.

In summary, one can draw parallels between the high-soft/low-hard classification for XRBs and the observed states of AGN, although there are many differences. For example, AGN do not cycle between the high-soft/low-hard states during disk instability-driven outbursts like XRBs, and the mechanism that causes AGN to perform this transition (with its associated jet quenching/launch) is unknown. There may be a correlation between SMBH spin and jet power, although this is not observationally confirmed.

With this background in the jet-disk connection and the factors that control the launching of a SMBH jet, we now turn towards VT J0243. We consider two scenarios. First, VT J0243 may be a young jet launched from a system that has been actively accreting since long before the jet was launched, i.e., an AGN. Alternatively, VT J0243 may be a jet launched near the onset of accretion. In this case, much of the previous discussion must be altered, as the properties of very young accretion disks are distinct from old disks (in particular, the magnetizations). The combination of young accretion and a new jet is expected for TDEs, so we discuss the possibility that VT J0243 is a jetted TDE.

\subsubsection{VT J0243 as a young jet from an AGN}

First, we consider the possibility that VT J0243 is a young jet from an AGN. We briefly compare the observations to the theory summarized in the previous subsections, and then we perform a detailed comparison of the observations of VT J0243 and known, young AGN jets.

From a theoretical perspective, even if all of the X-ray emission is due to an accretion disk/corona, VT J0243's bolometric luminosity is sufficiently low that it is feasible that we are observing an AGN in the low-hard state that has launched a jet. The lack of dust, based on the infrared colors and X-ray absorption,  and the low luminosities inferred from the IR and optical observations support the hypothesis that any pre-existing accretion disk was in a low state. 
The low average bulk Lorentz factor of the outflow ($\Gamma < 1.3$) is also consistent with the $\Gamma < 2$ jets typically associated with this state.

Of course, we cannot exclude that this event had an Eddington ratio ${\gtrsim}0.1$ during the jet launching, although the infrared observations and X-ray limits constrain the Eddington ratio to $\lesssim 1$. If the Eddington ratio is ${\gtrsim}0.1$ but not near- or super-Eddington, VT J0243 is in a regime where the physics of jet activity is very unclear. As we have discussed, in XRBs these higher Eddington ratios are associated with no jet activity. However, AGN in this regime are observed to be radio loud, and the mechanism through which the radio-emitting jet is produced is not fully understood (see \citealp{Liska2021} for simulations of a thin accretion disk that can support jet activity).

VT J0243 is consistent with theoretical expectations, albeit with uncertainties due to the unknown Eddington ratio at the time of jet launch. To further constrain the origin of VT J0243, we compare its properties with past observations. First, we compare the properties of VT J0243 and its host to the population of {\it persistent} radio-loud Seyferts. Later, we will focus back to transient sources and young jets. 

Astronomers have discovered jetted Seyfert galaxies, like VT J0243, although they are uncommon. Around ${\sim}15\%$ of broad line AGN are very radio-loud, where radio loudness is measured by the parameter $R_{\rm RL} = f_{\rm 6\,cm}/f_{\rm 4400\,\AA}$ and $R_{\rm RL}>100$ is the cut for very radio-loud AGN \citep[][]{Komossa2006}. In contrast, only ${\sim}2.5\%$ of Seyfert 1s have $R_{\rm RL}>100$, so these galaxies tend to be radio quiet \citep[][]{Komossa2006}. Radio loud Seyferts may have high black hole masses ${\sim}10^{7-8}\,M_\odot$ compared to the general Seyfert population, but still much lower masses than general radio-loud AGN (${\sim}10^9\,M_\odot$) \citep[][]{Komossa2006}. These black hole masses for radio-loud Seyferts are still higher than observed for VT J0243. Radio-loud Seyferts also have flat X-ray spectra ($\Gamma \sim 2$ for radio-loud Seyferts compared to $\Gamma \sim 2.9$ for the general Seyfert population) with rapid variability on as short as hour timescales \citep[][]{Komossa2018}. Note that the typical X-ray spectral slopes of radio-loud Seyferts are shallower than that of VT J0243. Radio-loud Seyferts have high Eddington ratios, and show strong Fe\,II emission, both in contrast. Finally, ${\sim}70\%$ of radio-loud Seyferts show compact, steep radio SEDs, analogous to the more general compact, steep spectrum (CSS) source population. This compact emission suggests an overabundance of young radio-emitting jets, which do not form into ${\sim}$kpc scale structures like observed in FR I/II galaxies \citep[][]{Berton2020}. In summary, the population of persistent radio-loud Seyferts shows some similarities to VT J0243, but many distinctions.

VT J0243 is not a persistent source, of course. Candidate young radio jets in AGN and Seyferts have become more common in recent years. \cite{Mooley2016} reported an AGN that switched from radio-quiet to radio-loud on a decade timescale, and more recently, \cite{Kunert2020} and \cite{Wolowska2021} published the first samples of such objects. We show individual light curves for these turning-on radio AGN in Figure~\ref{fig:lum_sum}. VT J0243 has a luminosity and timescale consistent with these events. 

Likewise, VT J0243 is consistent with observations of the jet power and bolometric luminosity of young, radio-loud AGN, which occupy specific regions of jet power$-$bolometric luminosity parameter space \citep[][]{Wolowska2021}. Adopting $P_J = 5\times10^{22} (L_{\rm 1.4\,GHz}/{\rm W\,Hz^{-1}})^{6/7}$ erg s$^{-1}$ \citep[][]{Rusinek2017}, and using the X-ray luminosity to approximate the bolometric luminosity with a bolometric correction factor ${\sim}20$ \citep[][]{Lusso2012}, we find $P_J \sim 10^{43.2}$ erg s$^{-1}$ and $L_{\rm bol} \sim 10^{43.6}$ erg s$^{-1} \sim 0.046L_{\rm edd}$ for VT J0243. This low Eddington ratio places the source slightly above the border of the radiatively inefficient regime, where most of the AGN energy is channeled into a radio-emitting jet. This regime is typically defined as $L_{\rm bol}/L_{\rm edd.} \lesssim 10^{-2}$. Given the large uncertainties in the bolometric luminosity of VT J0243, as discussed previously, we cannot convincingly place VT J0243 on either side of this dividing line. If we adopt $L_{\rm bol}/L_{\rm edd.} \sim 10^{-1.3}$ and $P_J/L_{\rm bol} \sim 10^{-0.4}$, we find VT J0243 is consistent with radio-detected AGN \citep[][]{Wolowska2021}.

On the other hand, VT J0243 has a unique radio SED relative to typical young jetted AGN. Young radio jets from AGN are observed to fall on a characteristic line in peak frequency$-$linear size parameter space \citep[e.g.][]{Nyland2020}. VT J0243 has a significantly smaller linear size compared to other young radio-loud AGN with the same peak frequency, which are typically hundreds of parsec in size.

Even if we only consider Seyferts, VT J0243 has unusual radio SED properties. A few expamples of bright radio flares from Seyferts have been detected. \cite{Lahteenmaki2018} observed 66 radio-quiet, narrow line Seyfert 1 galaxies at 37 GHz, and detected eight. These sources were undetected in archival observations from the VLA Faint Images of the Radio Sky at Twenty-Centimeters (FIRST) survey. They show variability at 37 GHz as large as a Jansky and on month$-$year timescales. Seyferts can produce bright radio flares. In contrast to VT J0243, the low frequency emission from these Seyferts is weak (${\sim}$micro-milliJy), suggestive of strong absorption at low frequencies \citep{Berton2020}. 

In summary, while VT J0243 may be consistent with expectations for a young jetted AGN from a theoretical perspective, its radio SED is distinct from typical young jetted AGN, it has a soft X-ray spectrum, and its host properties are unusual. For example, it has quiescent host IR colors, a lack of strong evidence for ongoing AGN activity within a few thousand years of the radio flare, and a low black hole mass. Given the large range of properties of young jetted AGN and the large theoretical uncertainties, we do not rule out that we are observing such an event. However, if VT J0243 is a young jetted AGN, it is an extremely unusual member of this class.


\subsubsection{VT J0243 as a young jet from a TDE} \label{sec:tde}

Tidal disruption events (TDEs) occur when a star ventures within the tidal radius, $R_T \sim R_* (M_*/M_{\rm BH})^{1/3}$, of a nearby SMBH \citep[e.g.][]{Rees1976, Rees1988,vanVelzen2011,Donley2002}. The bulk of TDEs are ``thermal'' TDEs with $\nu L_{\nu, {\rm GHz}}\lesssim 10^{38}$ erg s$^{-1}$, which are dominated by a thermally-emitting, hot accretion disk in the soft X-ray, and its reprocessed emission at lower energies \citep{Alexander2020}. The radio emission mechanism for thermal TDEs is poorly constrained, but may be associated with a disk wind or stellar debris outflow that is colliding with the CNM \citep{Alexander2020}. As is clear from Figure~\ref{fig:lum_sum}, VT J0243 is much brighter than all known thermal TDEs. 

The luminosity of VT J0243 is, however, consistent with the jetted TDE population, which includes the three brightest ($\nu L_{\nu, {\rm GHz}}\gtrsim 10^{40}$ erg s$^{-1}$) of the ${\sim}20$ radio-detected TDEs \citep[][]{Cenko2012, Pasham2015, Brown2017, Zauderer2011, Berger2012, Wiersema2020, Zauderer2013, Yang2016, Eftekhari2018}. The radio properties of these events are best exemplified through Sw J1644+57, the earliest example of an on-axis, jetted TDE. Sw J1644+57 was discovered by the {\it Swift} Burst Alert Telescope in 2011, and was promptly observed by a variety of telescopes across the electromagnetic spectrum. Within a few days, a radio outflow was detected at a luminosity near $10^{40}$ erg s$^{-1}$ and best-modelled as relativistic ($\Gamma \sim 3$) with $\nu_{\rm sa}<\nu_m$. The energy in the outflow increased over ${\sim}300$ days from ${\sim}2\times10^{50}$ erg to ${\sim}4\times 10^{51}$ erg while $\Gamma$ decreased as $\sim t^{-0.2}$. ${\gtrsim}300$ days post-launch, the energy plateaued, the peak flux began decreasing, and the SED transitioned to the regime with $\nu_{\rm m}<\nu_{\rm sa}$. The outflow transitioned to non-relativistic motion ${\sim}700$ days post-launch. 

Around the same time as the radio turned-on, Sw J1644+57 exhibited a bright X-ray flare peaking at an isotropic luminosity ${\sim}10^{48}$ erg s$^{-1}$, which is ${\sim}2-3$ orders of magnitude brighter than the Eddington luminosity of the SMBH. The X-ray emission declined as ${\sim}t^{-5/3}$, corresponding to the mass fallback rate during a TDE, and showed strong variability on $<1$day timescales. At $500$ days post-launch, the X-ray emission plummeted precipitously to $L_{\rm X}\sim10^{36}$ erg s$^{-1}$, which has been interpreted as the jet turning off. Because the X-ray luminosity tracks the expected mass fall back rate after a TDE, it is thought to be powered by a mechanism closely tied to the jet. \cite{Crumley2016} comprehensively surveyed many possible mechanisms, and favored models in which the X-ray photons are produced through either synchrotron emission or inverse-Comptonization of external photons (i.e., off the accretion disk). The emitting electrons are likely accelerated by magnetic reconnection in a Poynting flux-dominated jet. In this case, the fact that the observed jet was on-axis allowed the X-ray emission to be beamed, enabling the extremeley high luminosities observed.

In contrast to Sw J1644+57, and other similar events, VT J0243 is not associated with hugely super-Eddington X-ray flare. Only one other jetted TDE candidate was not detected as a bright X-ray transient, and this event was off-axis and in the highly obscured nucleus of a merging galaxy. VT J0243 may also be an off-axis jetted TDE. If we assume VT J0243 is a jetted TDE, this suggests that there will be a population of such events that cannot be detected via, e.g., X-ray transient surveys, but require wide field, deep radio surveys like VLASS.

VT J0243 also differs from Sw J1644+57-like events in its radio lightcurve. The ${\sim}5$ GHz luminosity is still increasing $>1000$ days post-launch, whereas ``typical'' jetted TDEs have long since begun fading at similar frequencies. Moreover, at $>1000$ days post-launch the SED is still in the regime where $\nu_{\rm sa} < \nu_m$. These observations may suggest that the jet has yet to turn off. Unusually slowly evolving TDEs are not unprecedented: some observed non-jetted TDEs evolve on much slower timescales than expected (see \citealp{Somalwar2021} Section 7.1 and references therein). The timescale of a TDE depends on factors including the stellar orbital parameters, the stellar structure, and the energy dissipation rate of the tidal debris. We may be seeing the jetted analogue of events in a regime with, e.g., a low energy dissipation rate, such that the accretion disk formation is delayed and the evolution slowed. In the case of VT J0243, the jet launch may have been enabled by magnetization provided by a fossil accretion disk, as was proposed in the case of Sw J1644+57 \citep[][]{Tchekhovskoy2014}.

In summary, VT J0243 is plausibly a jetted TDE. However, it differs from known jetted TDEs because of the slow timescale of the radio evolution and lack of an X-ray counterpart, the latter of which may support the idea that we are observing an off-axis jet.

\section{Conclusion} \label{sec:conc}

We have presented an extraordinarily bright and long lasting radio flare in a galactic nucleus detected in the VLA Sky Survey. VT J0243 rose to ${\sim}10^{40}$ erg s$^{-1}$ in a time period of ${\sim}5-20$ years. Radio follow-up suggests the presence of a compact, relativistic jet. X-ray emission with a luminosity $L_X = 10^{42.3}\,{\rm erg\,s}^{-1}$ is observed, which may be associated with a pre-existing or transient corona and accretion disk. Faint IR variability and an $L_g = 10^{42}$ erg s$^{-1}$ optical flare are observed, both consistent with reprocessed emission from a sub-Eddington EUV flare. VT J0243 is hosted by a weak Seyfert galaxy. A more detailed summary of our observations is provided at the beginning of Section~\ref{sec:disc}. 

VT J0243 is a unique example of a young radio source. It is likely caused by the launch of a powerful jet, combined with strongly sub-Eddington multiwavelength flares. This is consistent with a tidal disruption event, although the TDE likely evolved very slowly. It may also be an AGN, but the trigger for the abrupt accretion enhancement is unknown. In either case, VT J0243 highlights the complicated connection between SMBH accretion and jet launching. In the near future, radio surveys like VLASS will hopefully uncover large populations of similar, nascent jets, which, combined with extensive multiwavelength follow-up as was performed in this work, will illuminate the true triggers of such dramatic radio flaring and their connection with SMBH activity.

\acknowledgments

We would like to thank Amy Lien for her help with the Swift BAT data analysis.

Based on observations obtained with XMM-Newton, an ESA science mission with instruments and contributions directly funded by ESA Member States and NASA. The Australia Telescope Compact Array is part of the Australia Telescope National Facility (grid.421683.a) which is funded by the Australian Government for operation as a National Facility managed by CSIRO. We acknowledge the Gomeroi people as the traditional owners of the Observatory site. Basic research in radio astronomy at the Naval Research Laboratory is funded by 6.1 Base funding. This research has made use of MAXI data provided by RIKEN, JAXA and the MAXI team. The Dunlap Institute is funded through an endowment established by the David Dunlap family and the University of Toronto. B.M.G. acknowledges the support of the Natural Sciences and Engineering Research Council of Canada (NSERC) through grants RGPIN-2015-05948 and RGPIN-2022-03163, and of the Canada Research Chairs program. P.C. acknowledges support of the Department of Atomic Energy, Government of India, under the project no. 12-R\&D-TFR-5.02- 0700. We thank the staff of the GMRT that made these observations possible. The GMRT is run by the National Centre for Radio Astrophysics of the Tata Institute of Fundamental Research. The National Radio Astronomy Observatory is a facility of the National Science Foundation operated under cooperative agreement by Associated Universities, Inc. We thank the staff of the GMRT that made these observations possible. GMRT is run by the National Centre for Radio Astrophysics of the Tata Institute of Fundamental Research. Some of the data presented herein were obtained at the W. M. Keck Observatory, which is operated as a scientific partnership among the California Institute of Technology, the University of California and the National Aeronautics and Space Administration. The Observatory was made possible by the generous financial support of the W. M. Keck Foundation. The authors wish to recognize and acknowledge the very significant cultural role and reverence that the summit of Maunakea has always had within the indigenous Hawaiian community.  We are most fortunate to have the opportunity to conduct observations from this mountain. This work made use of data supplied by the UK Swift Science Data Centre at the University of Leicester.

\bibliography{J0243-2840}{}
\bibliographystyle{aasjournal}

\end{document}